\documentclass[useAMS,usenatbib,usegraphicx]{mn2e}
\usepackage{amsmath,amssymb}
\usepackage{xcolor}

\newif\ifAMStwofonts
\AMStwofontstrue

\newcommand{\simlt}{\lower.5ex\hbox{$\; \buildrel < \over \sim \;$}}
\newcommand{\simgt}{\lower.5ex\hbox{$\; \buildrel > \over \sim \;$}}

\title[Dust extinction law in MW]
{Constraints on the dust extinction law of the Galaxy with {\sl Swift}/UVOT, {\sl Gaia}
and {\sl 2MASS}}
\author[I. Ferreras et al.]
{Ignacio Ferreras$^{1,2,3,4}$\thanks{E-mail: i.ferreras@ucl.ac.uk}, 
M\'onica Tress$^2$, Gustavo Bruzual$^5$, St\'ephane Charlot$^6$,\and
Mat Page$^2$, Vladimir Yershov$^2$, Paul Kuin$^2$, Daisuke Kawata$^2$,
Mark Cropper$^2$\\
$^1$ Instituto de Astrof{\'i}sica de Canarias, Calle V{\'i}a L{\'a}ctea s/n,
E38205, La Laguna, Tenerife, Spain\\
$^2$ Mullard Space Science Laboratory, University College London,
Holmbury St Mary, Dorking, Surrey RH5 6NT, UK\\
$^3$ Department of Physics and Astronomy, University College London, London WC1E 6BT, UK\\
$^4$ Departamento de Astrof\'\i sica, Universidad de La Laguna, E38206 La Laguna, Tenerife, Spain\\
$^5$ Instituto de Radioastronom\'\i a y Astrof\'\i sica, UNAM, Campus Morelia, Michoac\'an, M\'exico C.P. 58089, M\'exico\\
$^6$ Sorbonne Universit\'e, CNRS, UMR7095, Institut d'Astrophysique de Paris, F-75014 Paris, France
}

\begin{document}
\date{MNRAS, in press. Accepted 2021 April 30. Received 2021 April 30; in original form 2020 November 17}
\pagerange{\pageref{firstpage}--\pageref{lastpage}} \pubyear{2021}
\maketitle
\label{firstpage}


\begin{abstract}
We explore variations of the dust extinction law of the Milky Way by
selecting stars from the {\sl Swift}/UVOT Serendipitous Source
Catalogue, cross-matched with {\sl Gaia} DR2 and {\sl 2MASS} to
produce a sample of 10,452 stars out to $\sim$4\,kpc with photometry covering a
wide spectral window.  The near ultraviolet passbands optimally
encompass the 2,175\AA\ bump, so that we can simultaneously fit the
net extinction, quoted in the $V$ band (A$_V$), the steepness of the
wavelength dependence ($\delta$) and the bump strength (E$_b$). The
methodology compares the observed magnitudes with theoretical stellar
atmospheres from the models of Coelho. Significant correlations are
found between these parameters, related to variations in dust
composition, that are complementary to similar scaling relations found
in the more complex dust attenuation law of galaxies -- that also
depend on the distribution of dust among the stellar populations
within the galaxy.  We recover the strong anticorrelation between
A$_V$ and Galactic latitude, as well as a weaker bump strength at
higher extinction. $\delta$ is also found to correlate with latitude,
with steeper laws towards the Galactic plane. Our results suggest that
variations in the attenuation law of galaxies cannot be fully
explained by dust geometry.
\end{abstract} 


\section{Introduction}
\label{Sec:Intro}

Dust is one of the key observable components of galaxies. Through
absorption and scattering, dust affects the way starlight reaches the
observer, and a good characterization of the wavelength dependence of
these processes is fundamental to be able to, for instance, derive
accurate stellar masses or star formation rates of galaxies. There are
two main ways in which the presence of dust will shape the 
spectral energy distribution of a galaxy: the dust properties will lead to an
extinction law that reflects how the superposition of dust grains of
different size and composition absorbs and scatters incoming
light \citep[see, e.g.][]{Draine:03}. In addition, the distribution
of dust within the galaxy will affect light from the constituent stellar
populations in different ways. Most importantly, light from other regions
of the galaxy may be scattered  {\sl into} the line of sight to the observer,
complicating the interpretation of attenuation, that no longer depends
exclusively on dust composition \citep[see, e.g.][]{WG:00,Panuzzo:07}.
A deeper understanding of the dust extinction law in the Milky Way Galaxy and
its variations, will help remove these degeneracies.
Note studies of dust extinction in the Milky Way 
are free from such effects, as we are simply probing direct lines of sight to
stars. We emphasize that extinction and attenuation are different concepts
measured in similar ways.
For instance, clumpiness in the distribution of dust, typically 
found in star forming regions, will affect the steepness of the
wavelength dependence of the attenuation
law \citep[e.g.,][]{WG:00}, whereas the same steepness parameter
only depends on dust composition when considering extinction,
as it is measured along a single line of
sight towards a star \citep[e.g.,][]{FM:90}. This paper focuses solely
on dust extinction, as we are observing single stars in our
Galaxy. Therefore, any variations found in the derived parameters can only be
caused by changes in dust composition.
\\
The observations in the Milky
Way and the Magellanic Clouds reveal an overarching trend towards higher
extinction at shorter wavelengths, along with 
a reduced set of spectral features, most notably the 2,175\AA\ NUV
bump and the 9.7$\mu$m and 18$\mu$m silicate features
\citep[see, e.g.][]{Galliano:18}.
The most likely candidate to explain the NUV bump -- targeted in this
paper -- is related to the resonance present in benzene-based
structures such as Polycyclic Aromatic Hydrocarbons \citep[PAHs, see, e.g.][]{DS:98}.
However there are other
potential carriers \citep{Bradley:05}, so a full account of this
feature remains an open question.
There has been a number of parameterizations of the wavelength dependence
of the extinction law in the Milky Way, 
and the observational constraints led to 
the definition of standard extinction laws, although with a
large amount of scatter among different sight lines
\citep[see, e.g.][]{CCM:89, FM:90,Fitz:99}.
The extinction
law of the Small Magellanic Cloud \citep[SMC,][]{Pei:92,Gordon:03} features 
a steeper wavelength dependence, characteristic of smaller dust
grains \citep[see, e.g.,][]{Wein:01}.
In addition, the SMC law also lacks the NUV bump
\citep{Hagen:17}.
Knowledge of the dependence of the
scatter of the extinction law on dust composition allows us to extend
the analysis to dust attenuation in unresolved galaxies to find out
about dust composition \citep[see, e.g.][]{Tress:19}

The UVOT camera \citep{Roming:05} on board the {\sl Neil Gehrels Swift
Observatory} features three NUV filters, UVW2, UVM2 and UVW1, that
straddle the 2,175\AA\ bump, along with three standard
$U$, $B$, $V$ filters in the optical spectral window.
The optimal location of the NUV passbands has been exploited in studies
to constrain the dust attenuation law in nearby
galaxies \citep[e.g.,][]{Hoversten:11,Hutton:14,Decleir:19} as well
as the dust extinction law in the SMC \citep{Hagen:17}. For instance, 
in \citet{Hutton:14} it was shown that the attenuation law in
nearby prototypical starburst galaxy M82, cannot be explained by
the standard starburst prescription of \citet{Calz:00}, as the NUV bump
is present throughout the galaxy. Moreover, a substantial radial
gradient was evident in the strength of the bump and the
total-to-selective extinction parameter, R$_V$ \citep{Hutton:15}.
\\
This paper explores a large archival sample of stars observed by the
UVOT camera, cross-matched with the {\sl Gaia} and {\sl 2MASS} catalogues, as presented
in \S\ref{Sec:Data}. The methodology that allows us to extract
constraints on the dust extinction parameters is presented in
\S\ref{Sec:Method}, including an analysis based on simulated data to
explore the systematics. A discussion of the results is shown in Sec.~\ref{Sec:Disc},
followed by a summary in \S\ref{Sec:Summ}. Throughout this paper, 
all magnitudes are quoted in the AB system \citep{ABmag}.

\begin{figure}
\begin{center}
\includegraphics[width=88mm]{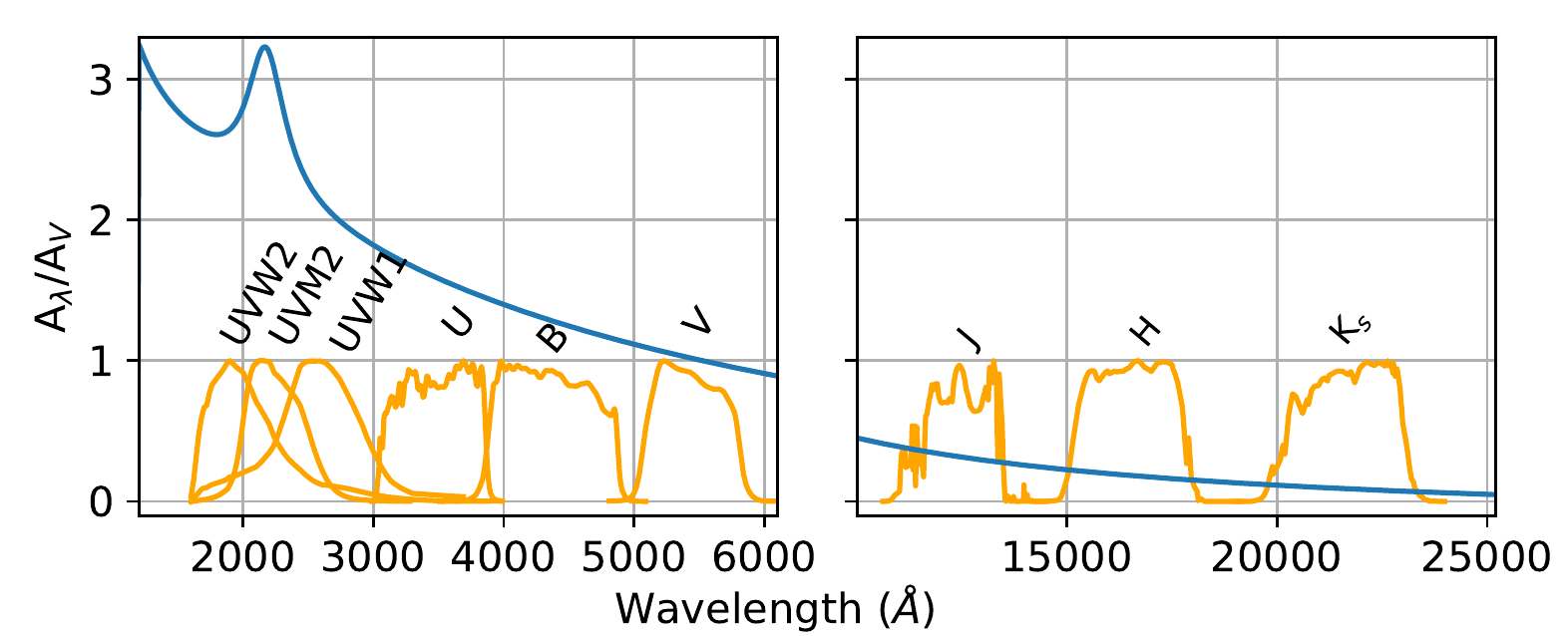}
\end{center}
\caption{The standard extinction law usually adopted for the Milky Way
(corresponding to $E_b=4$ and $\delta=-0.05$) is shown, along with
the normalized response of the passbands used in the analysis.}
\label{fig:Extinc}
\end{figure}

\begin{figure}
\begin{center}
\includegraphics[width=85mm]{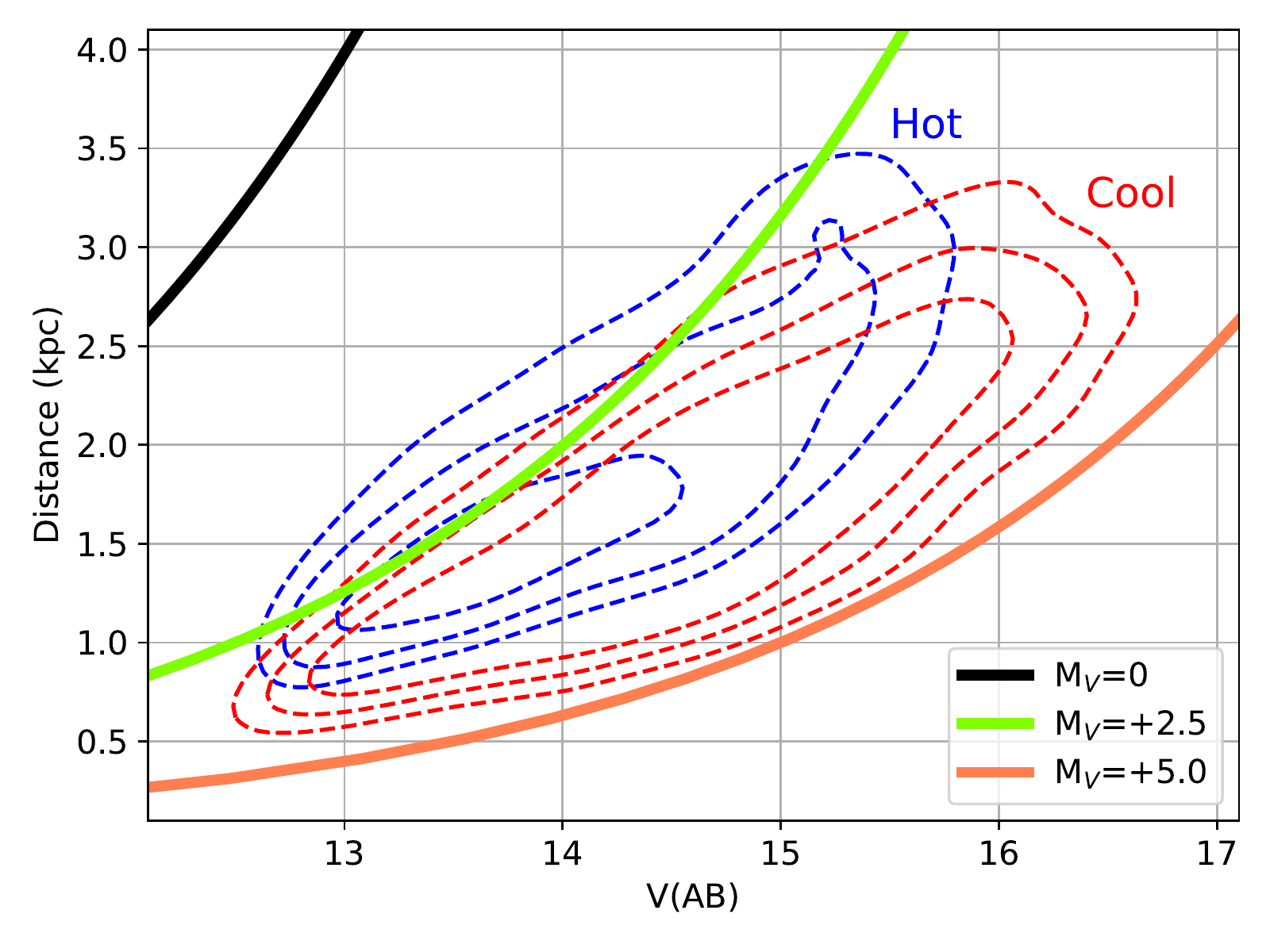}
\end{center}
\caption{Distribution of stars on the distance vs $V$ band
apparent magnitude plane.
The contours trace the density of stars of our working sample on this
diagram, split with respect to effective temperature at the median
value (T$_{\rm eff}$=7,500\,K). From the inside out, the contours
encompass 90, 95 and 99\% of the hot/cool subsets. For reference, the black, green and
orange lines represent the expected loci of an unreddened star with
absolute $V$ band magnitude of 0, $+$2.5 and $+$5, respectively.}
\label{fig:DvsV}
\end{figure}

\section{Data}
\label{Sec:Data}

This work focuses on photometric data from the
UVOT Serendipitous Source catalogue \citep[UVOT/SSC,][]{UVOTCat1,UVOTCat2},
a compilation of all sources (resolved and unresolved)
found in the archive of observations taken by
the {\sl Swift}/UVOT instrument, produced by processing all archival data.
We use v1.1 of the UVOT/SSC catalogue, comprising a total of
23,059 observations in which over 13.86 million sources are detected,
although we note that the same sources could have been observed repeatedly in
several runs.  The number of unique sources in this catalogue, with
available fluxes in all six photometric bands is 170,801.  From this
set, we select those observations with good photometric measurements
-- defined as those with an uncertainty below 0.2\,mag in all six
passbands, along with a quality flag equal to zero. This constraint
ensures good accuracy in the derivation of the dust extinction parameters,
as we show below, and only decreases the sample size by 8.8\%.
The sample is
cross-matched with the {\sl Gaia} DR2 \citep{GDR2} stellar catalogue, selecting
those sources with an available effective temperature
from \citet{GDR2_Teff}, and a distance estimate
from \citet{GDR2_Dist}. The set is also matched with the point
source catalogue of {\sl 2MASS} \citep{2MASS} to include NIR photometry
in the $J$, $H$ and $K_s$ passbands. The cross-matching was done
with the help of the excellent service provided by the
Centre de Donn\'ees astronomiques de Strasbourg
(CDS)\footnote{\tt http://cdsxmatch.u-strasbg.fr}.
The sample from this cross-match comprises 80,104 stars.
Fig.~\ref{fig:Extinc} shows the response of the nine passbands
used in this study -- separated into the UVOT NUV and optical filters,
(left) and the 2MASS NIR filters (right). For reference, the standard
dust extinction law of the Milky Way is shown \citep{CCM:89}.

One of the key parameters explored in this paper is the 2,175\AA\  bump,
that requires a significant amount of flux in the NUV bands. Trial
and error in our methodology (see below) results in setting a lower
limit in the effective temperature to be able to constrain the
dust extinction parameters. We opted to impose a threshold in the
effective temperature, T$_{\rm eff}>6,500$\,K, producing a final working
sample that comprises 10,452 stars. We emphasize that the temperature
mentioned here has been obtained independently, using a supervised machine
learning algorithm \citep{GDR2_Teff}, trained on a standard assumption
for the dust extinction law. Since we adopt a more generic function,
the actual values of the temperature may vary. However, we only use
this one to impose a Gaussian prior in the analysis, as shown in the
next section. Hence, the final estimates of temperature will be
consistent with the best fit extinction law for each star.

The distribution of the sample in the distance versus apparent
magnitude plane is shown in Fig.~\ref{fig:DvsV}, split at the median
T$_{\rm eff}$ (7,500\,K) with hot and cold stars shown as blue/red
dashed contours, respectively.  From the inside out, the contours
engulf 90, 95 and 99\% of each subset.
We note that due to both sample selection and the
actual distribution of stars as a function of temperature,
the cooler subsample mostly consists of A- and F-type stars.
For reference, the figure includes the expected location of
unreddened stars with absolute $V$-band magnitude of 0, +2.5 and +5.0 (black,
green, and orange lines, respectively). The number of stars drops at
apparent magnitudes fainter than $V\simgt$16.5\,AB, and extends to a distance
$\simlt$4\,kpc. It is also worth noting that a segregation based on colour
does not introduce a bias regarding distance or apparent
magnitude. This will be relevant when we explore in \S\ref{Sec:Disc} 
the distribution of dust extinction parameters.

\begin{figure}
\begin{center}
\includegraphics[width=85mm]{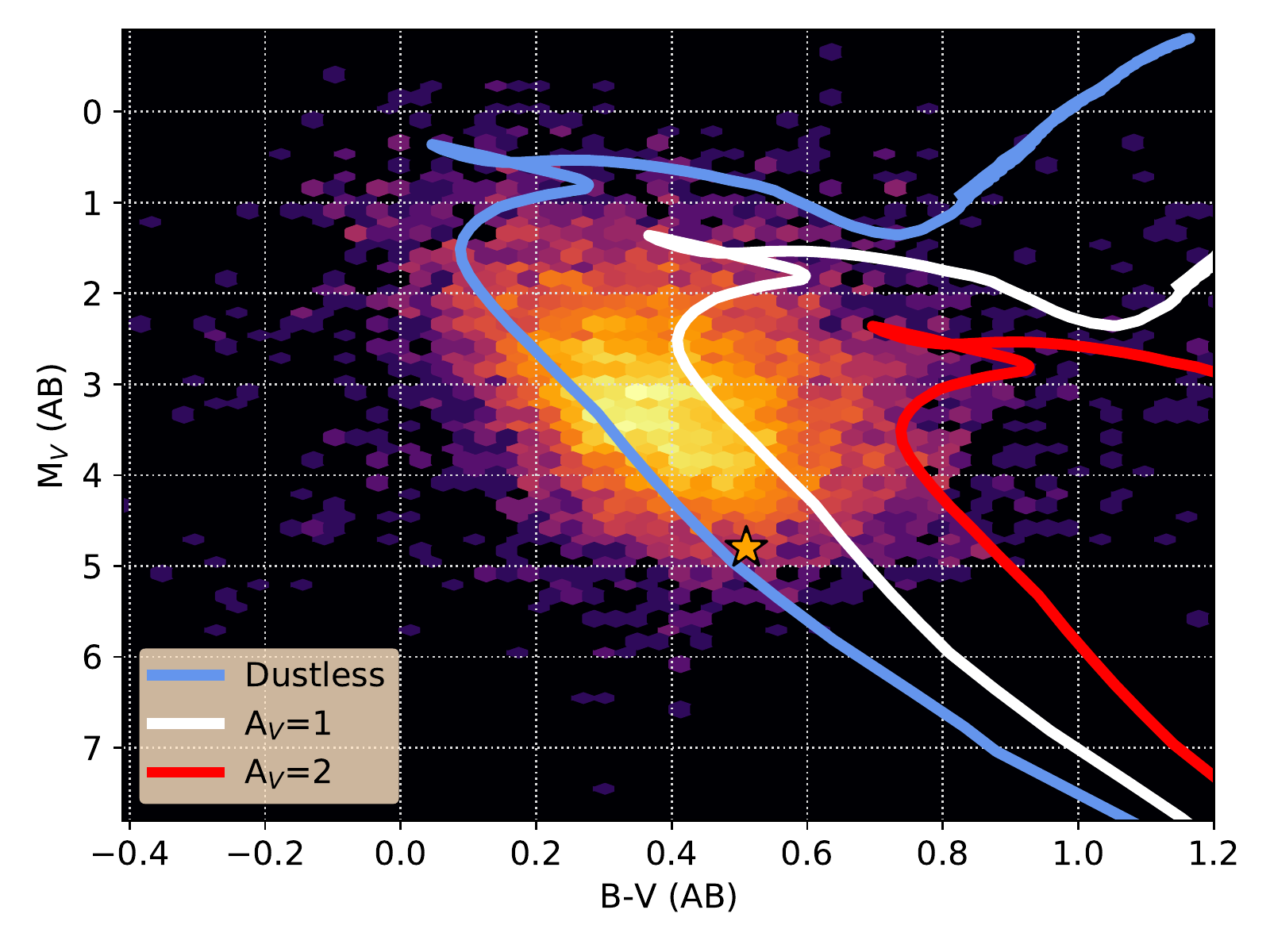}
\end{center}
\caption{Distribution of the sample on the absolute $V$ band magnitude vs $B-V$
plane. The conversion from apparent to absolute magnitude makes use of
the {\sl Gaia} DR2 distances. For reference, a solar metallicity
isochrone at 1\,Gyr of age (from the PARSEC team, \citealt{Bressan:12},
see, e.g., \citealt{Alzate:21}) is shown in the dustless case (blue),
or introducing a $V$ band extinction of A$_V$=1 (white) or 2\,mag (red) for
a Milky Way standard dust law \citep{CCM:89}. The location of
the Sun is shown in this diagram as a star, for reference.}
\label{fig:Isoch}
\end{figure}

Fig.~\ref{fig:Isoch} shows the sample in a Hertzsprung-Russell diagram, 
with $V$-band absolute magnitude against $B-V$ colour. For reference, a
solar metallicity, 1\,Gyr isochrone from the PARSEC team
\citep{Bressan:12} is overlaid, including the effect of extinction, as
labelled, adopting a standard Milky Way extinction
law \citep[$R_V$=3.1, see, e.g.][]{CCM:89}. Models based on these
isochrones provide accurate fits to the observed colour-magnitude
distribution of the {\sl Gaia} data \citep{Alzate:21}.  This choice is
only meant to illustrate the location of our data on a standard
colour-magnitude diagram. As the population ages, the Main Sequence
turnoff will shift towards redder and fainter values, in the same
direction as an increased dust extinction, reflecting the well-known
age-dust degeneracy of stellar populations. This work does not attempt
to fit isochrones, and focuses instead on a combined constraint of the
temperature and dust extinction from the observed fluxes of single
stars. This figure illustrates the nature of our sample as mostly
turn-off and sub-giant stars with a range of extinction
A$_V\simlt$2\,mag. Our sample is biased in
favour of the brighter and bluer sources, as we impose detection in
all NUV and optical filters from the UVOT catalogue as the main
selection criterion. This implies a preponderance of stars on the Galactic
plane (90\% of the sample has Galactic latitude $b\lesssim 20^{\rm o}$),
and allows us to cover a wide range of distance, with 90\% of the sample
probing out to $\lesssim 3.3$\,kpc, without a significant bias regarding
temperature (Fig.~\ref{fig:DvsV}).

\begin{table}
\caption{Range of Stellar and Dust Parameters\label{tab:params}}
\begin{center}
\begin{tabular}{ccc}
\hline
Parameter & Range & Gridpoints\\
\hline
\multicolumn{3}{c}{Stellar parameters (log\,g=4.5, cgs)}\\
\hline
T$_{\rm eff}$ & [4000, 18000]\,K  & 42 \\
$[{\rm Fe/H}]$ & $\{-0.5,0,+0.2\}$ & 3\\
\hline
\multicolumn{3}{c}{Dust parameter (flat) priors$^a$}\\
\hline
$A_V$ & $[0,8.5]$ & ---\\
$E_b$ & $[0,11]$ & ---\\
$\delta$ & $[-1,+1]$ & ---\\
\hline
\end{tabular}
\end{center}
$^a$ Note the parameters are explored as a continuous
function with an MCMC sampler, and the stellar spectra
are retrieved from a bilinear (T$_{\rm eff}$, [Fe/H]) interpolation
of the model grid.
\end{table}

\begin{figure*}
\begin{center}
\includegraphics[width=140mm]{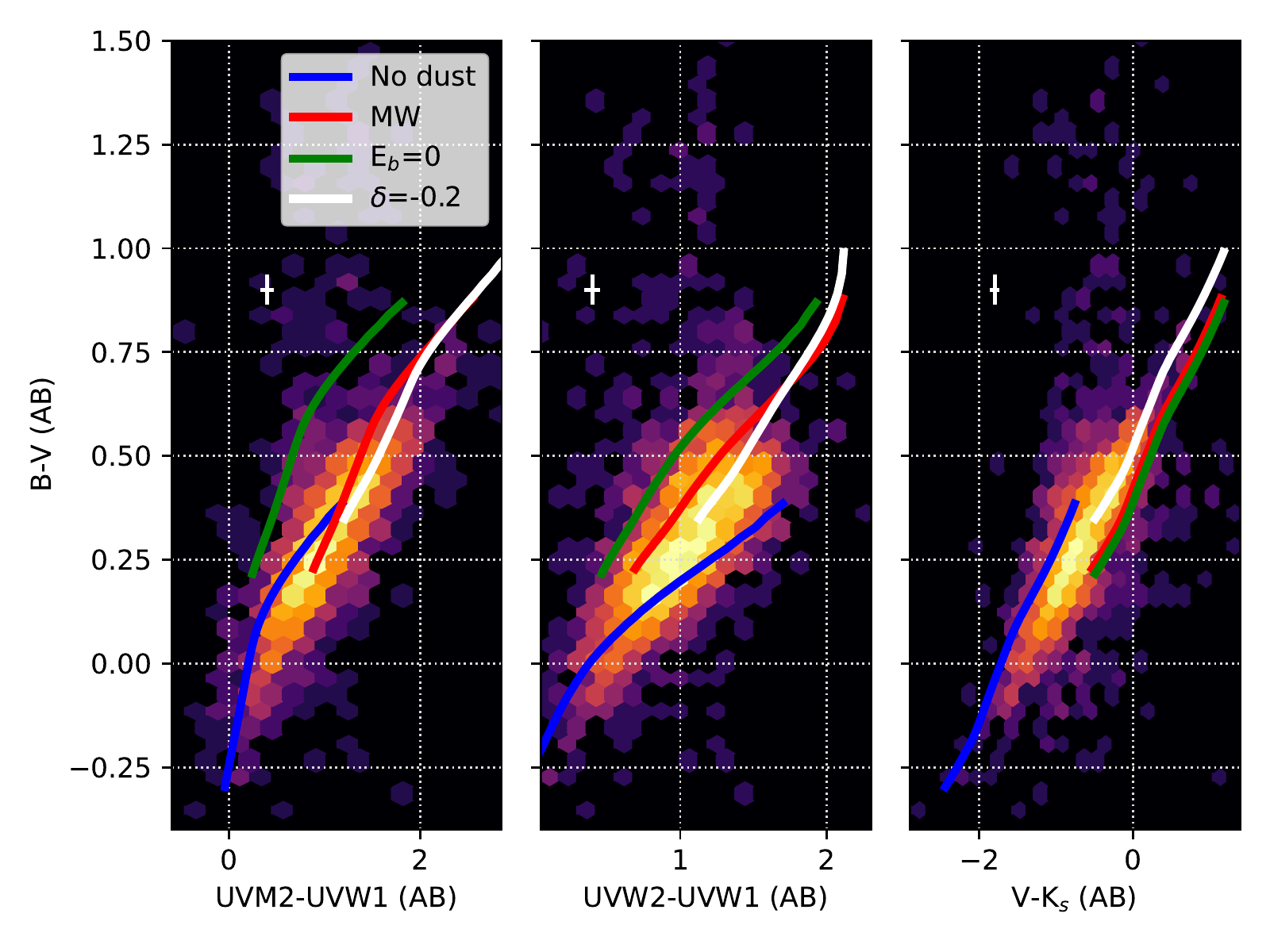}
\end{center}
\caption{Colour-colour diagrams of the sample.
The panels show the optical $B-V$ 
plotted against two NUV colours from UVOT
and an optical$-$NIR colour ($V-K_s$).
The blue lines are tracks from the stellar library used in this
paper, at solar metallicity over a range of effective
temperatures (6,500-18,000\,K), without any dust correction applied.
The red lines show the effect of a Milky Way standard dust correction
($E_b=4$, $\delta=-0.05$), at $A_V=2$, and the green lines are the equivalent case, with the
NUV bump set to zero ($E_b=0$, $\delta=-0.05$). The white lines adopt a change in the
steepness of the extinction law, towards a stronger wavelength dependence
($\delta=-0.2$ corresponds to R$_V\sim$2.6, keeping the original bump strength $E_b=4$).
A 1\,$\sigma$ error bar in each panel illustrates 
the worst case scenario for the statistical uncertainty of the colour estimates.
}\label{fig:CCD}
\end{figure*}

\begin{figure}
\begin{center}
\includegraphics[width=85mm]{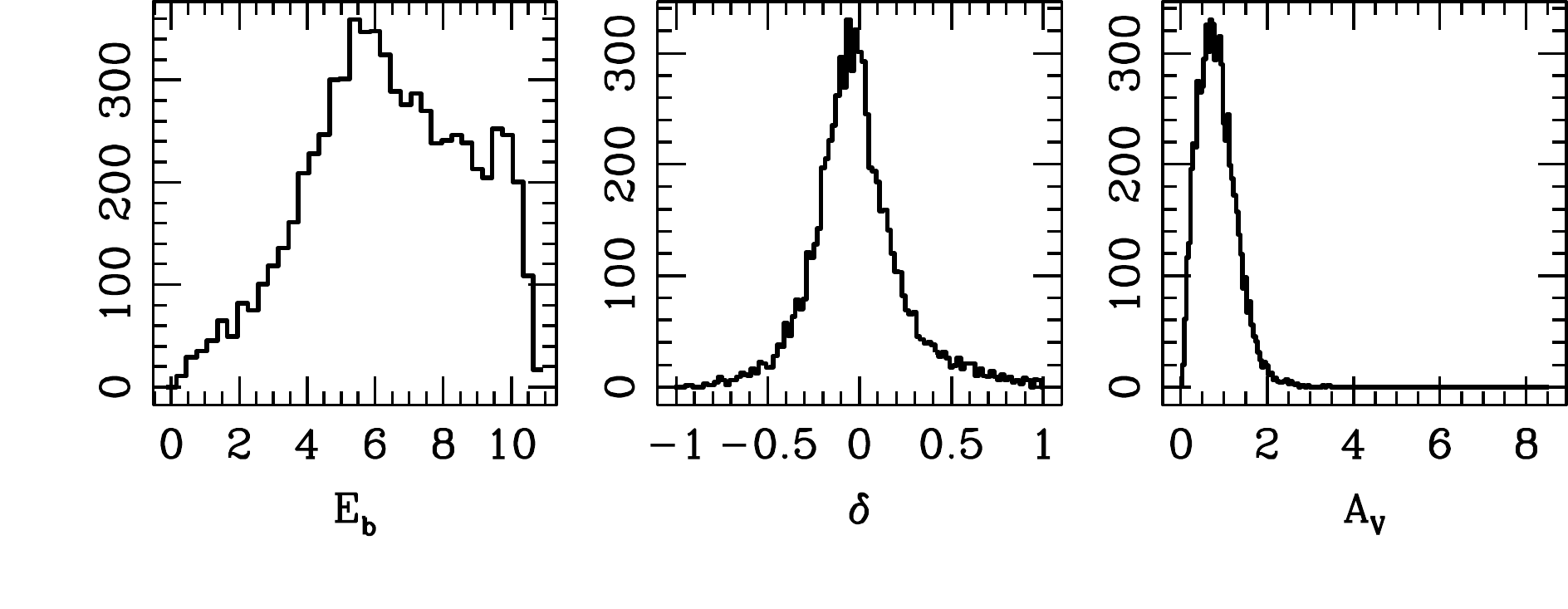}
\end{center}
\caption{Distribution of the best-fit dust extinction parameters.
From left to right, the bump strength ($E_b$), steepness parameter ($\delta$) and
$V$-band extinction ($A_V$) are shown.}
\label{fig:hist}
\end{figure}

\section{Methodology}
\label{Sec:Method}

There are several functional forms in the literature to describe the
wavelength dependence of dust extinction and attenuation \citep[see,
e.g.][]{CCM:89,FM:90,FM:07,Noll:09,CSB10}.  We choose the parametric function  
presented by \cite{KC:13} from the previous work of
\citet{Noll:09}\footnote{There is no evident advantage of using this parameterisation
with respect to other choices, but we find this one is used rather
extensively in the literature.}.
At each wavelength, $\lambda$, the extinction, in magnitudes,
can be written:
\begin{equation}
A(\lambda)=\frac{A_V}{4.05}\left[k^\prime(\lambda) + E_bD(\lambda)\right]
\left(\frac{\lambda}{\lambda_V}\right)^\delta,
\label{eq:ExtLaw}
\end{equation}
where $k^\prime(\lambda)$ is the standard reddening curve
found in equation~4 of \citet{Calz:00}. 
The NUV bump is described by a Lorentzian-like profile, 
$D(\lambda)$ (often termed a Drude profile), scaled by the NUV bump strength parameter, $E_b$,
with two additional parameters, the central wavelength and the
width. We follow \citet{KC:13} to define this absorption feature, 
namely adopting 2,175\AA\ for the central wavelength and 350\AA\ for the width. 
The central 
wavelength does not vary among different lines of sight, but the
width does show significant scatter \citep[see, e.g.,][]{Valencic:04,FM:07}. However, it is beyond the
scope of this paper to treat the width as a free parameter, so that
our constraints of $E_b$  should be taken as the effective
bump strength for a fixed absorption profile.
Finally, $\delta$ is a parameter that changes the steepness of the original
wavelength dependence of the \citet{Calz:00}
law  -- that corresponds to $E_b$=0, $\delta$=0. Positive/negative values
of $\delta$ correspond to a shallower/steeper wavelength dependence, respectively.
Note that the number 4.05 in equation~\ref{eq:ExtLaw} refers to the original total to selective
extinction parameter R$_V$ for the Calzetti law. Therefore, the wavelength
dependence of this prescription is concisely defined
by three parameters: A$_V$, E$_b$ and $\delta$.
An alternative description of the extinction/attenuation law 
is the one defined by \citet{CSB10}, where the steepness of the law
is parameterised by the more traditional $R_V$, and
the bump strength is described by $B$, so that the choice
R$_V$=3.1, B=1 closely follows the standard 
Milky Way law of \citet{CCM:89}. In the appendix
of \cite{Tress:18} a simple analytic expression is presented to
convert between these two parameterisations. For instance, the
Milky Way standard corresponds to $\delta\simeq -0.05$, E$_b\simeq 4$.
We emphasize that our analysis, based on broadband photometry, does not
consider independent variations in the wavelength dependence in the NUV
and optical windows, a task that can only be attempted by use of spectroscopic data
\citep[e.g.][]{FM:90,Gordon:09,Fitz:19}.

\begin{figure}
\begin{center}
\includegraphics[width=85mm]{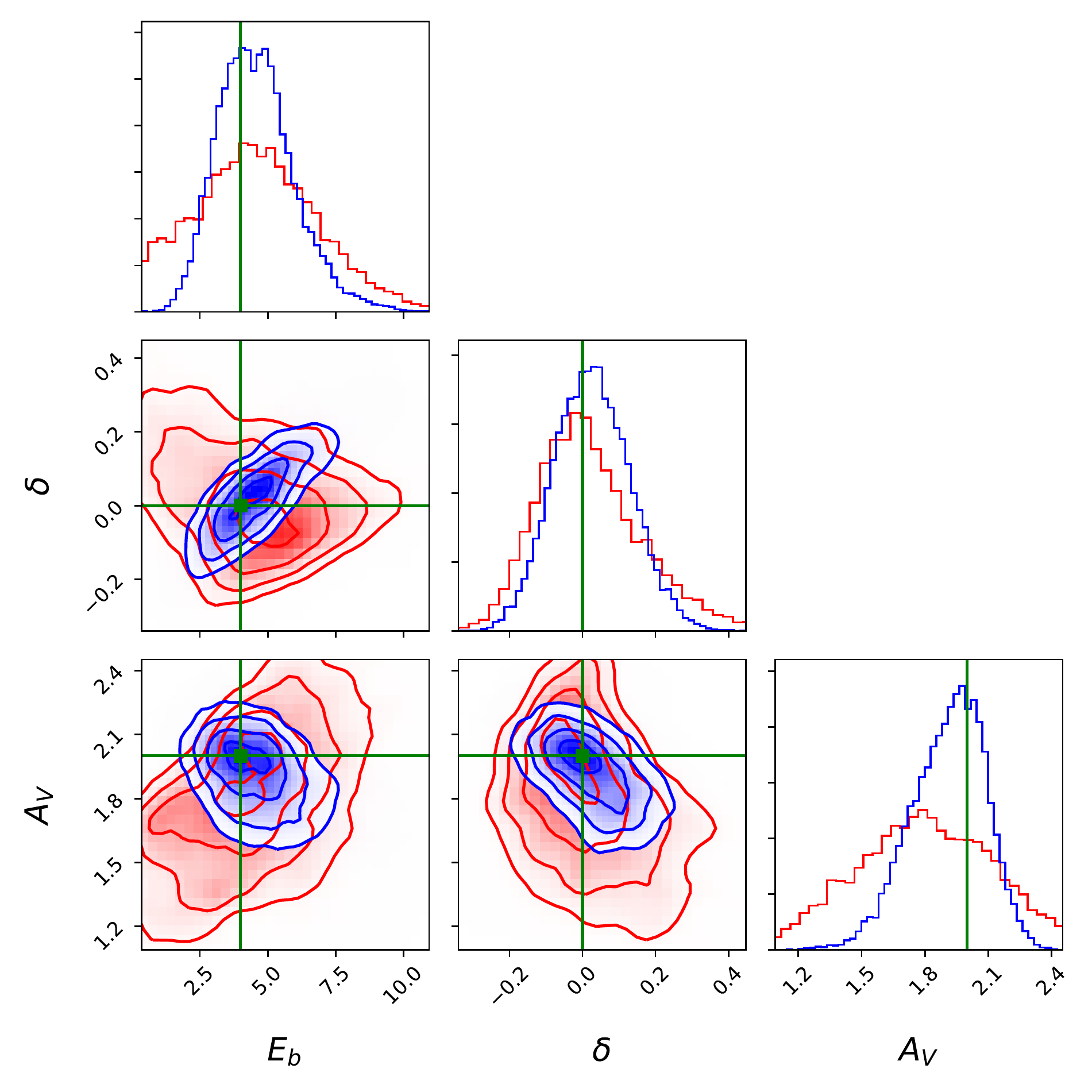}
\end{center}
\caption{Retrieval of input parameters in a mock observation with 10\% uncertainty
in the fluxes, with dust extinction parameters E$_b$=4, $\delta$=0, A$_V$=2.
Two stars are considered, both
at solar metallicity. The four contours delimit the posterior probability at the
0.5, 1, 1.5 and 2$\sigma$ confidence levels (from the inside out).
In red we show a T$_{\rm eff}$=6,500K star and in blue
a hotter T$_{\rm eff}$=10,000K star.
The true values of the dust parameters are shown by the green straight lines and square
symbol. The retrieved parameters are  E$_b$= 4.47$\pm$2.28, $\delta$=$+$0.01$\pm$0.15,
A$_V$= 1.80$\pm$0.33 for the cooler star and 
E$_b$= 4.47$\pm$1.42, $\delta$=$+$0.02$\pm$0.10, A$_V$= 1.93$\pm$0.18 for
the hotter star (1\,$\sigma$ uncertainties). The maximum likelihood is significantly different, with the
hotter star peaking at ${\cal L}$=0.92 whereas the cooler star
only reaches ${\cal L}$=0.14.
This figure illustrates the higher accuracy at hotter effective temperatures,
especially concerning the NUV bump strength.}
\label{fig:MCMC}
\end{figure}

Our methodology rests on a comparison between the combined
NUV, optical and NIR photometry with a set of stellar atmospheres
from the models of \citet{Coelho:14}. The spectra are retrieved in {\sc FITS} format
from Coelho's webpage\footnote{\tt http://www.astro.iag.usp.br/$\sim$pcoelho/}.
These models include a homogeneous
computation of opacity distribution functions, and provide a set of
stellar spectra for a wide range of effective temperature, surface
gravity and chemical composition, including non-solar abundance ratios.
However, since our analysis is based on low spectral resolution photometry
using traditional broadband filters, we restrict the parameter space to
effective temperature and metallicity, keeping a fixed value of surface
gravity and abundance ratios. We adopt these models as they have been thoroughly
tested in population synthesis models by \citet{Coelho:20}, where it is shown that 
some of the existing discrepancies in the output of synthetic populations
based on either empirical or theoretical stellar
libraries is mostly due to the sampling of the underlying stellar parameters, rather than the
stellar spectra themselves. Therefore,  spectra from
state-of-the-art stellar atmosphere models are suitable for this
type of analysis and free from the dust correction residuals
present in empirical data. Moreover, the computed colours based on 
filters with broad bandpasses only carry systematic uncertainties that are 
substantially smaller than the typical
statistical uncertainties of the observations. We emphasize here the
need to use theoretical libraries, and note that some of the stars in
empirical libraries that cover NUV wavelengths, such as
NGSL \citep{NGSL} do suffer from a prominent absorption feature at the
position of the NUV bump. This absorption is caused by dust along the
line of sight to the star, that would require an accurate knowledge of
the properties of dust in the Milky Way along specific
directions. Therefore, the model predictions from empirical spectra will
be compromised.

\begin{figure}
\begin{center}
\includegraphics[width=85mm]{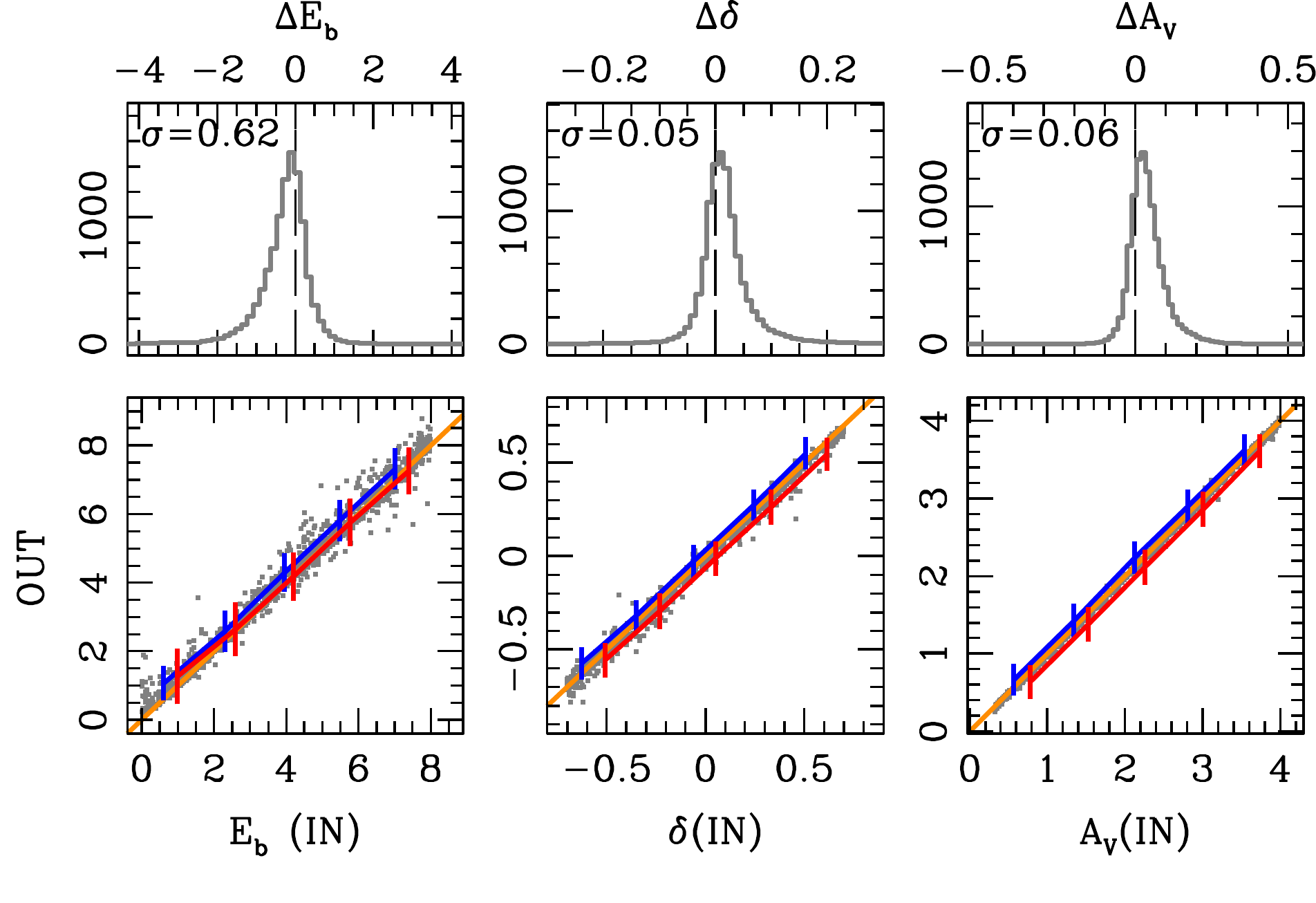}
\end{center}
\caption{Retrieval of dust extinction parameters from mock observations
produced from the theoretical stellar spectra subject to dust
extinction obtained from a random selection of the parameters. The
bottom row shows the comparison between the input parameters
(horizontal) and those determined by the MCMC algorithm (vertical),
with the orange line marking the ideal 1:1 correspondence. The
individual measurements are shown as grey dots.  The blue and red lines
show the averaged results in the two subsamples -- split at the median
T$_{\rm eff}$=7,500\,K -- with the higher/lower effective
temperature in blue and red, respectively.  The error bars represent
the 1\,$\sigma$ scatter. The top row shows the distributions $\Delta
p\equiv p_{\rm IN}-p_{\rm OUT}$, where $p$ is one of the three dust
extinction parameters (from left to right: E$_b$, $\delta$, A$_V$).
The number in each panel gives the standard deviation.  }
\label{fig:Sims}
\end{figure}

To illustrate the power of a photometric analysis to constrain the
dust extinction law, we show in Fig.~\ref{fig:CCD} three colour-colour
diagrams with our sample represented as a density plot. Three tracks
are overlaid that correspond to models, at solar metallicity, solar
abundance ratios and surface gravity log\,g=4.5 (cgs).  The tracks
cover the temperature range from T=6,500\,K (top-right corner of the
three panels) to 18,000\,K, and are shown at zero dust extinction
(blue), for a Milky-Way standard law (i.e. $\delta=-0.05$, $E_b=4$)
with A$_V=2$\,mag (red), and for the same amount of extinction with a
law that lacks the NUV bump (represented by $E_b=0$, keeping
$\delta=-0.05$, green).  The white line keeps the same bump strength
as that of the Milky Way ($E_b=4$), but changes $\delta$ to a steeper
value ($-0.2$) to show the effect of a change in the slope of the
extinction law. The error bars show the worst case scenario for the
statistical uncertainty of the colour measurements. Note, for
instance, the lack of sensitivity to the bump strength of $V-K_s$, as
expected, whereas the NUV colours show a substantial
sensitivity. Although UVW2$-$UVW1 avoids the central part of this
resonance, it does show a dependence on E$_b$ given the rather
extended width of this feature. Variations in the slope produce
changes in colour that progressively decrease with wavelength. Making
use of all these different sensitivities to the dust parameters,
combining photometric information from NUV to NIR along with the
optimal spectral range of the UVOT NUV filters, it is possible to
constrain all three extinction parameters.

For each star in the cross-matched UVOT sample, we compare the observed fluxes with a range of stellar
atmosphere models based on a grid of 42 values of effective
temperature, from T$_{\rm eff}$=4,000\,K to 18,000\,K and three values
of metallicity: [Fe/H]={$-$0.5,0.0,$+$0.2}. We emphasize that although we do
have a previous estimate of T$_{\rm eff}$ from \citet{GDR2_Teff}, the value is
not consistent with a generic dust extinction law as the one adopted here. Therefore,
it is expected that, due to the temperature-dust degeneracy, the effective
temperature may change either to lower or higher values. Therefore, we take
stellar atmosphere models with lower temperatures than the threshold of 6,500\,K.
At the other end, we limit the highest temperature in the models
to 18,000\,K given the drop in the number of stars shown in the
colour-colour plots at the hot end of the isochrone trails (Fig.~\ref{fig:CCD}).
Also note that in this sample, none of the values of T$_{\rm eff}$ from the analysis
of \citet{GDR2_Teff} reaches temperatures above 10,000\,K.
Moreover, the fits to the observed stars produce a temperature range markedly
lower, as shown below.
We seek a balance between runtime and accuracy. Given the low spectral
resolution of broadband photometry, we opted to restrict the stellar
parameters to a single value of surface gravity, log\,g=4.5 (cgs) and solar
abundance ratios ([$\alpha$/Fe]=0). Different values of surface
gravity do not vary significantly the colours based on broadband
fluxes. We emphasize that the dust parameters will only be constrained
with flux ratios (i.e. colours), so that the luminosity of a given
model cannot introduce any bias. For each trial, the solver chooses five
parameters, namely \{T$_{\rm eff}$, [Fe/H], A$_V$, $\delta$, E$_b$\}.
The first two are used to interpolate a stellar spectrum from the
grid, and the last three parameters allow us to attenuate the spectrum
using the adopted extinction law. For each observed star, we define a
standard (log-)likelihood:
\begin{equation}
\ln{\cal L} = {\cal A} - \frac{1}{2}\chi^2(T_{\rm eff},{\rm [Fe/H]},A_V,\delta,E_b),
\end{equation}
\noindent
where ${\cal A}$ is a normalization factor. 
For each choice of
the stellar parameters ($T_{\rm eff}$ and [Fe/H]), we extract
an interpolated spectrum from the grid of \citet{Coelho:14} models,
then the dust extinction corresponding to the choice of
dust parameters ($A_V$, $\delta$, $E_b$) is applied to the
spectra, and the magnitudes computed for each of the
filters, using the official passbands in each
case\footnote{The passband response curves can be found at the following link: {\tt http://svo2.cab.inta-csic.es/theory/fps/}}.
The value of the $\chi^2$ statistic is computed by comparing
observed and model magnitudes, taking the
V-band flux as an anchor point, meaning that all stars are
assumed to have the same $V$ band magnitude, so that only the
colours are used to constrain the parameters.
A standard MCMC sampler is applied to
explore the 5-dimensional parameter space, making use of the {\sc
Python} code {\sc emcee} \citep{emcee}.  The MCMC solver comprises 100
chains, taking 2000 steps, from which we remove the first 500 to
derive the probability distribution function (PDF) of each parameter.
The values of the parameters are quoted at the median of these
distributions.

\begin{figure}
\begin{center}
\includegraphics[width=85mm]{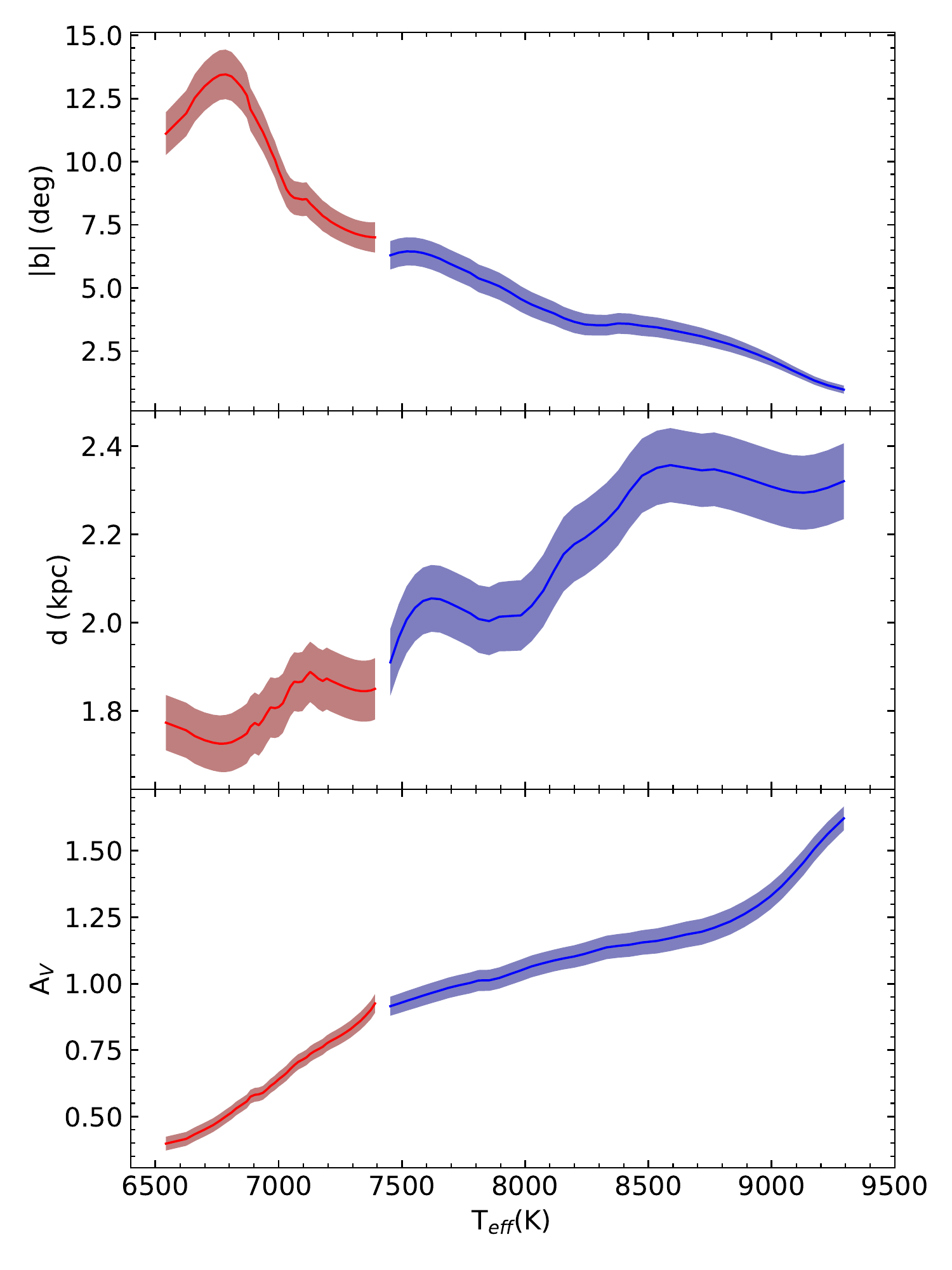}
\end{center}
\caption{The best fit effective temperature is compared with 
(from top to bottom) Galactic latitude, distance, and $V$ band
extinction. The shaded regions follow a running median along with its
uncertainty, binning 200 stars in each trial. The two colours represent
the split at the median T$_{\rm eff}$=7,500\,K, following the same
notation as in figures~\ref{fig:DPars} and \ref{fig:ParSky}.}
\label{fig:Teff}
\end{figure}

Flat priors are imposed on the dust parameters over a wide range of
values, see Table~\ref{tab:params} for details of the range allowed. To
overcome the potential degeneracy between temperature and the dust
parameters, we impose a Gaussian prior centered at the value of
T$_{\rm eff}$ obtained by \citet{GDR2_Teff} with a standard deviation
corresponding to twice the quoted error. This way we will allow for a
variable T$_{\rm eff}$, but will avoid fits that correspond to
very different temperatures. Fig.~\ref{fig:hist} shows the distribution
of the best fit parameters for the total sample. It is worth noting
that only $\sim$1.4\% of the sample has $V$-band extinction higher than 2\,mag.
In the next section we explore the
accuracy of the parameter retrieval using a simulated sample, and will
determine whether substantial degeneracies are present.

\subsection{Simulations}
\label{Ssec:Sims}

We test the methodology by constructing a set of mock photometric
measurements, choosing a random set of dust- and stellar-related
parameters, to produce the NUV, optical and NIR fluxes.  To illustrate
the retrieval process for a single star, Fig.~\ref{fig:MCMC} shows
cuts of the posterior probability distribution for two stars with the
same dust extinction parameters (E$_b$=4, $\delta$=0 A$_V$=2,
represented by the green lines) and different effective temperatures:
6,500\,K (red) and 10,000\,K (blue). The contours are shown at
four confidence levels: 0.5, 1, 1.5, and 2\,$\sigma$. The lower temperature case
produces a less accurate constraint of the parameters, as expected,
since cooler stars do not have enough NUV flux to probe in detail the
extinction law.

Each synthetic observation has compatible photometric uncertainties
with respect to the original sample: the mock set comprises the same
number of stars (10,452), and for each one we use the original
T$_{\rm eff}$ and flux uncertainties.  In this way, we guarantee that the
simulations mimick as much as possible the properties of the original
sample. A random value of the metallicity (from a uniform deviate in
log space, between [Fe/H]=$-$0.5 and $+$0.2) and dust parameters
(also uniform deviates, E$_b\in[0,8]$, $\delta\in[-0.7,+0.7]$, A$_V\in[0.3,4]$)
are chosen for each star, producing the fluxes according to the
theoretical models adopted in this work. Note that the range is
different from the grid models explored in Table~\ref{tab:params}.
The grid models are expected to extend over a significantly larger
range than the actual dust parameters of the sample, for robustness.
Fig.~\ref{fig:hist} shows that the parameters extracted from the actual
observations are contained within the intervals explored in the
simulated data, thus justifying this choice. We note the distribution
of E$_b$ appears truncated at high values, but the results presented
below do not show any pile up at high bump strengths in the trends,
so we do not anticipate any systematic effect from the chosen range of
values for E$_b$. The comparison between input
and output are shown in Fig.~\ref{fig:Sims}.  The bottom row compares
the input parameters (horizontal axes), with respect to the ones
retrieved by the MCMC-based methodology described above.  Individual
mock stars are shown as grey dots, and closely follow the ideal 1:1
trend (orange lines). To explore a potential systematic, we split the
sample into high (blue) and low (red) effective temperature, with the
cut at 7,500\,K (that corresponds to the median of the sample).  The
lines show the averaged trends, with the error bars spanning one
standard deviation. No significant segregation is found, within error
bars, although the cooler stars consistently produce lower values of
A$_V$ and $\delta$, with differences that are, nevertheless very
small. The top row shows the distribution of the difference between input and
output (grey histograms), i.e. $\Delta p\equiv p_{\rm IN}- p_{\rm
OUT}$, where $p$ is one of the three dust extinction parameters.
The behaviour is acceptable, with
standard deviations shown in the panels, namely $\Delta$E$_b$=0.62;
$\Delta\delta$=0.05; $\Delta$A$_V$=0.06.  The 1\,$\sigma$
uncertainties extracted from the posterior of the individual star
measurements have median values $\sigma(E_b)$=1.63;
$\sigma(\delta)$=0.11; $\sigma(A_V)$=0.25. Although temperature is
treated as a nuisance parameter, the simulations produce good results
with differences between the input values and the extracted ones of
$\Delta T_{\rm eff}=37\pm 78$\,K.

We also emphasize that for each star the dust law is determined from a
random set of parameters ($A_V$, $\delta$, $E_b$), so that no spurious 
correlation should be expected between them. This sanity
check allows us to assess whether there is any systematic from the
underlying degeneracies when comparing the results with actual
stars. Columns 2 and 3 of Table~\ref{tab:Pearson} show the Pearson
correlation coefficients of the input and output parameters for four
pairings. The data for the output includes the 1$\sigma$ scatter
derived from bootstrapping according to the retrieved error in each
parameter. Most importantly, only a weak covariance is found between
A$_V$ and both bump strength and effective temperature, and a weaker
trend is found in the random simulations between steepness and bump
strength. Nevertheless, we will show in the next section that the mild
correlation found here is statistically smaller than the one obtained
for the actual sample.  The results of these simulations allow us to
assess the limit of applicability of the models. We emphasize that, at
present, this is the best possible data to probe the NUV and optical
dust extinction law in the Milky Way, with a sample comprising over 10
thousand stars.

\begin{figure}
\begin{center}
\includegraphics[width=85mm]{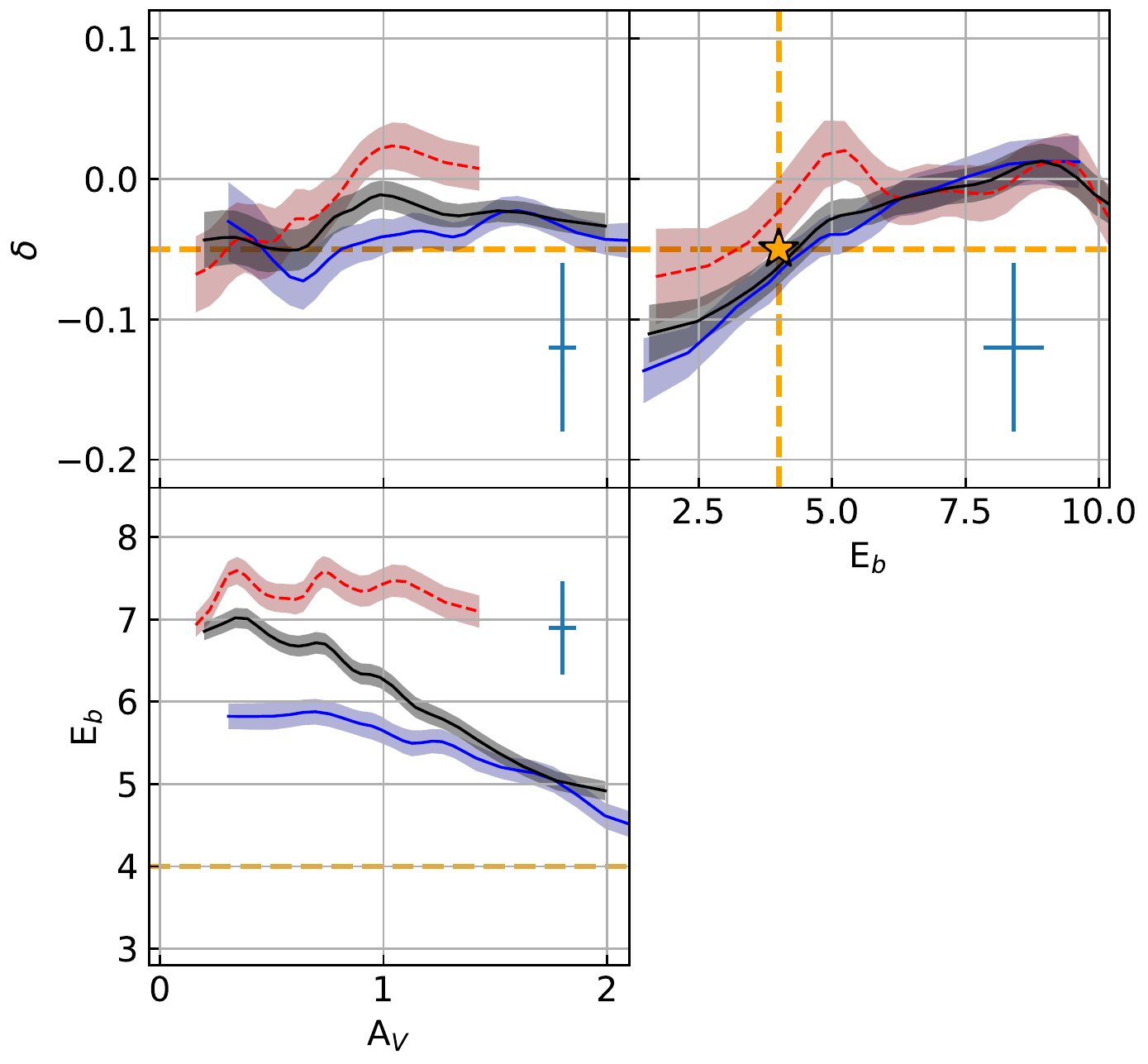}
\end{center}
\caption{The best-fit dust extinction parameters
are compared among one another.  In each panel, the lines represent a
running median with respect to the parameter shown in the abscissa,
with the shaded region encompassing the 1\,$\sigma$ error on the
median. Red and blue correspond to subsamples split with respect to
effective temperature, split at the median (7,500\,K) into a hot
(blue) and cold (red) subsample. The full sample is shown in grey. The
orange lines mark the dust extinction of the standard Milky Way law. The
median of the distribution of parameter uncertainties shown in each
panel as error bars.}
\label{fig:DPars}
\end{figure}

\begin{table}
\caption{Pearson correlation coefficients between parameters.
The values are obtained from a bootstrap that creates 200 random
realizations of the data, quoting the median and the standard
deviation in each case, from left to right: the input and output
from simulated data, used to test the methodology (see Sec.~\ref{Ssec:Sims}), 
and the actual observations.}
\label{tab:Pearson}
\begin{center}
\begin{tabular}{c|cc|c}
\hline
Params & \multicolumn{2}{c}{Mocks} & Observations\\
& IN & OUT & \\
(1) & (2) & (3) & (4)\\
\hline
$\delta$ vs $A_V$    & $-0.0144$ & $+0.0042\pm 0.0045$ & $+0.1105\pm 0.0058$\\
$E_b$ vs $A_V$       & $-0.0125$ & $+0.0415\pm 0.0063$ & $-0.2727\pm 0.0051$\\
$\delta$ vs $E_b$    & $-0.0065$ & $+0.0205\pm 0.0065$ & $+0.1032\pm 0.0073$\\
$A_V$ vs $T_{\rm eff}$ & $+0.0102$ & $+0.0460\pm 0.0081$ & $+0.5455\pm 0.0054$\\
\hline
\end{tabular}
\end{center}
\end{table}

\section{Results and Discussion}
\label{Sec:Disc}

\begin{figure*}
\begin{center}
\includegraphics[width=140mm]{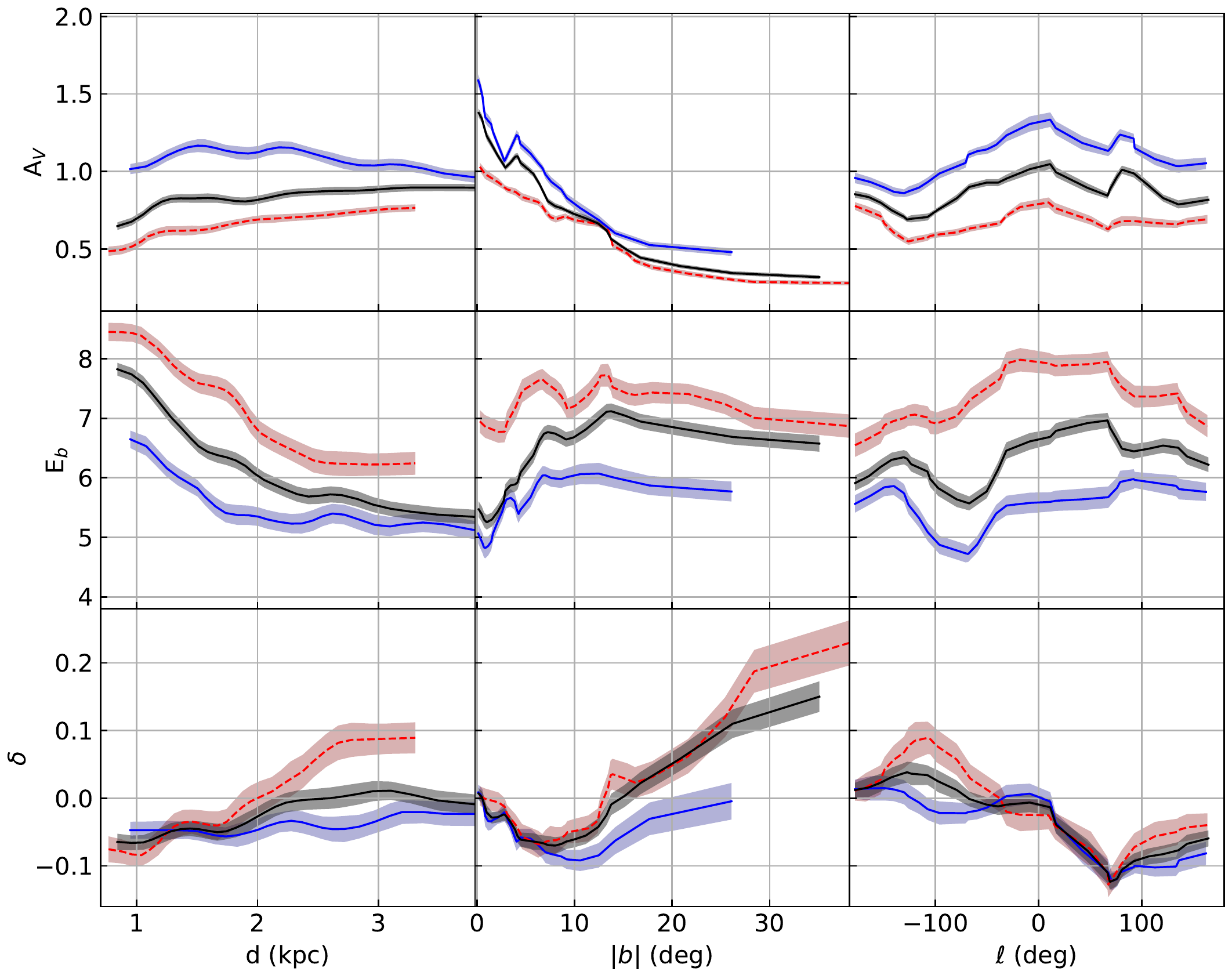}
\end{center}
\caption{Equivalent of Fig.~\ref{fig:DPars} for the
distribution of the dust extinction parameters
(ordinate, from top to bottom: $V$ band extinction,
NUV bump strength, and steepness of the wavelength dependence)
with respect to astrometric quantities (abscissa, from
left to right: distance, Galactic latitude, and longitude).
The plots show a running median for the whole sample (grey) and
split with respect to temperature into hotter (blue) and
cooler (red) stars, split at T$_{\rm eff}$=7,500\,K.}
\label{fig:ParSky}
\end{figure*}

Fig.~\ref{fig:Teff} shows the trends between the best fit effective
temperature and some of the parameters that define the sample, from
bottom to top, extinction in the $V$ band, distance and Galactic
latitude.  The shaded regions encompass a running median along with
its (1\,$\sigma$) uncertainty, using a moving interval that includes
200 stars in each trial.  In this plot, as well as in the next few
figures that apply a similar running median to show the trends of the
various parameters, please note that the range on the vertical axis
strongly depends on the choice of the horizontal axis, so that it
should not come as a surprise that these plots may extend over
different ranges, being always smaller than those of the original
parameter search of Table~\ref{tab:params}.  Regarding this plot,
T$_{\rm eff}$ is derived exclusively from the observed colours (we do
not use the absolute fluxes), therefore the trends found with distance
and latitude are non-trivial and strengthen the validity of our
approach, as we find a strong correlation towards higher temperature
stars with increasing distance (as expected in a flux limited survey)
and decreasing latitude, as expected from the well-known distribution
of stars on the Galactic plane \citep[e.g.][]{SG:07}.  The blue/red
shaded regions in this figure show the split with respect to effective
temperature at the median value (7,500\,K), as presented in
Figs.~\ref{fig:DPars} and \ref{fig:ParSky}, below.

The general comparison between the different 
parameters can be found in Fig.~\ref{fig:DPars}, where the three
extinction parameters are plotted against each other. We follow
the same split with respect to temperature as in Fig.~\ref{fig:Teff}.
For reference, the orange star and dashed lines mark the
Milky Way standard extinction law ($E_b=4$, $\delta=-0.05$), and a
characteristic error bar of the dust parameters is also shown in
each panel, defined as the median error bar for each parameter from
the MCMC-based fits to the whole sample, given at the $1\,\sigma$ level.
Regarding the $E_b$ parameter we note two main results:
1) There is a 
negative correlation between the NUV bump strength and
$V$-band extinction, mostly caused by the hotter subsample
that follows more closely the Galactic plane. 
The bump is stronger in the cooler subsample. However, this difference
might be partly caused by the fact that at lower temperature, the
stellar flux is rather weak in the NUV part of the spectrum, and the
bump is harder to constrain (see Fig.~\ref{fig:MCMC}), hence we put more
emphasis on the result for the hotter stars. 2)There is also 
a slightly positive correlation between $\delta$ 
and E$_b$, so that steeper laws (i.e. lower values of $\delta$) display weaker
bumps. This correlation is independent of the effective temperature.
However, the trend is rather mild, note the small variation of
the running median for $\delta$. 
These two results can be put in context with the equivalent
relations found in galaxies -- i.e. regarding the attenuation law,
and thus more complex to disentangle \citep{KC:13,Hagen:17,Salim:18,Tress:18,Decleir:19}.
The attenuation studies 
also find a similar trend towards decreasing $E_b$ with $A_V$, but the
correlation between $E_b$ and $\delta$ is opposite to the one found here
-- although the observational trends show a rather large scatter. We
emphasize that the trends with respect to attenuation also depend on
dust geometry, and, indeed, some numerical simulations can explain the
observed trends with models at constant composition but variable
dust geometry \citep{Nara:18}. These trends so far have only been
reported in dust attenuation studies in galaxies, not in the original
extinction curve. Our results suggest that substantial
variations in dust composition are present within the same galaxy
(in a relatively small volume, $d\lesssim 3$\,kpc), so that these
variations should be accounted for in any theoretical model of attenuation.
This result suggests the carrier responsible for the
NUV bump suffers significant variations across different lines of sight, with
a well-defined trend towards stronger bumps at low extinction and in
regions where the law is shallower. On a speculative
tone, one could argue that low extinction regions provide an environment
where the formation of large grains is less efficient (i.e. steeper wavelength
dependence, and thus, lower $\delta$), whereas in this environment the much
smaller carriers of the NUV bump are not destroyed, therefore leading to
higher E$_b$. At high values of A$_V$, the formation of larger grains and
a higher efficiency in the destruction of, e.g. PAHs, would lead to
the observed relations \citep{FD:11}. However, it is beyond the scope
of this paper to provide a physical scenario to explain these trends.

\begin{figure}
\begin{center}
\includegraphics[width=85mm]{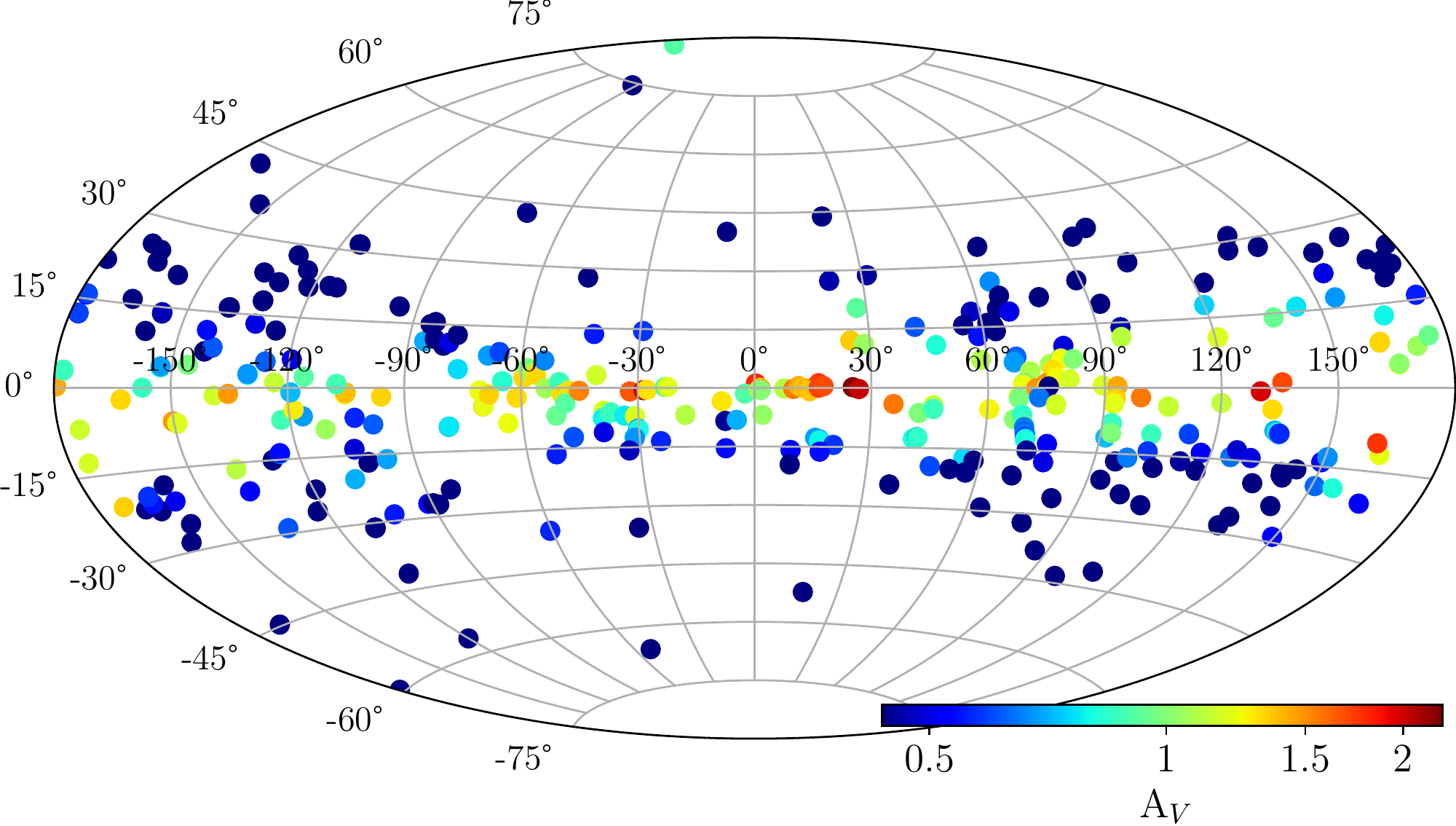}
\includegraphics[width=85mm]{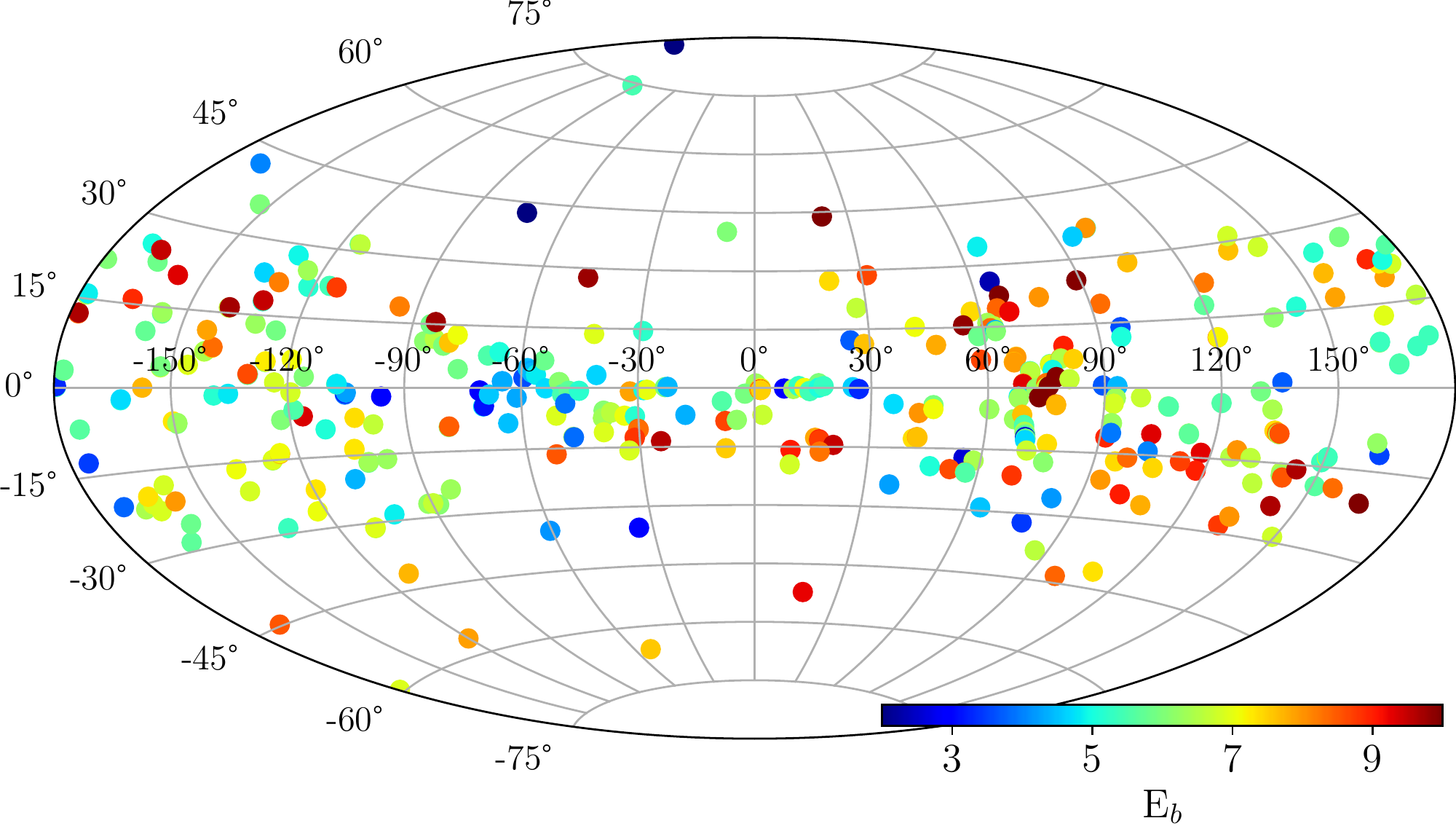}
\includegraphics[width=85mm]{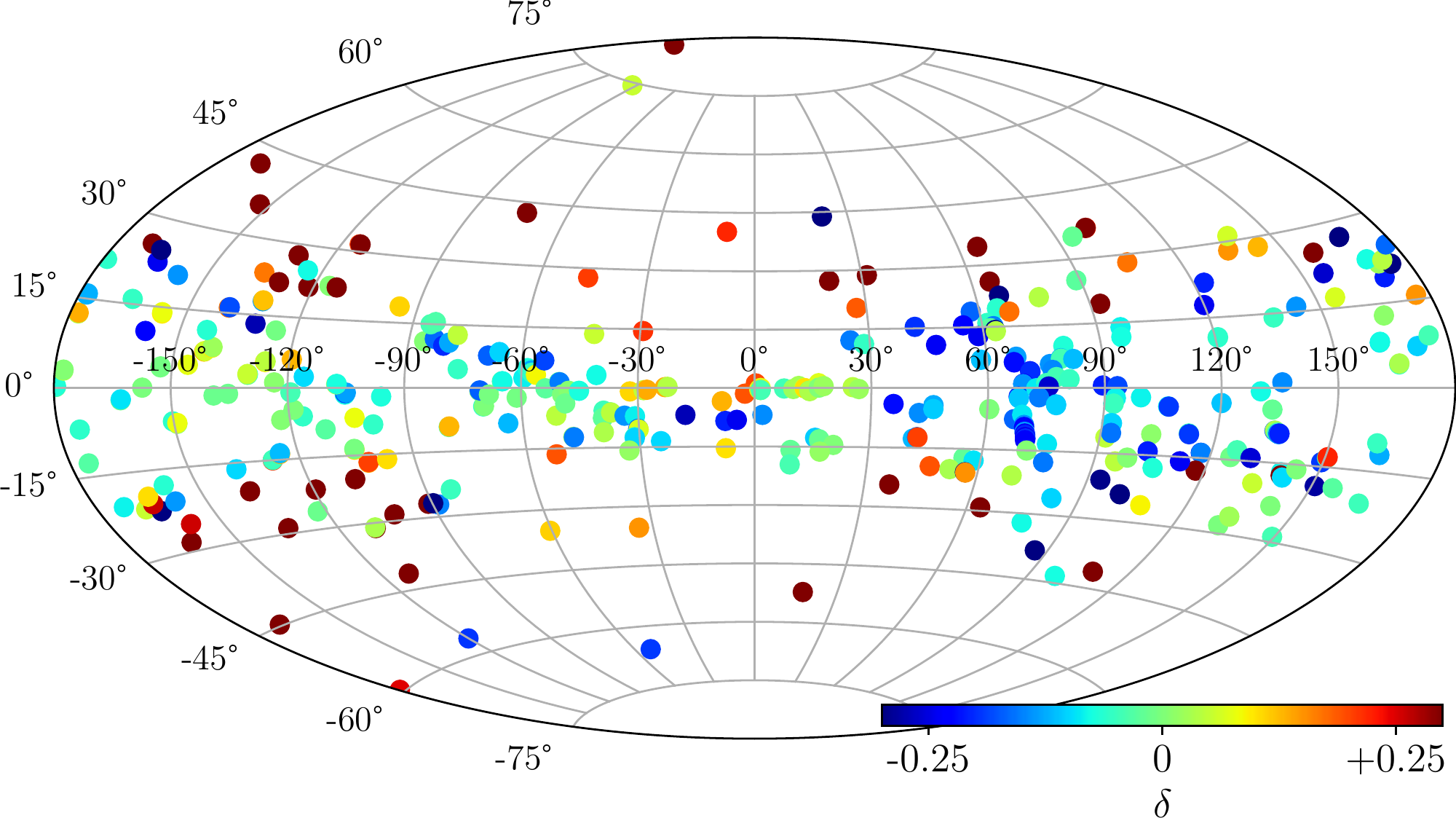}
\end{center}
\caption{Aitoff projections of our working sample, 
colour coded with respect to the three extinction parameters, from top to
bottom: $V$ band extinction (A$_V$), NUV bump strength
(E$_b$), and steepness of the wavelength dependence ($\delta$). Note that the
sample of 10,452 stars originates from the UVOT Serendipitous Source
Catalogue, that is not designed to map the whole sky. Therefore, the coverage
is sparse, and the whole set is reduced to pointings within the same 
field of view of the UVOT camera. The colour coded dots follow average
values within each pointing.}
\label{fig:Aitoff}
\end{figure}

The validity of these trends can also be quantified by comparing the
Pearson correlation coefficient of the data to those quoted above for
the best fits corresponding to mock simulations, whose parameters
were, by construction, uncorrelated. The results are shown in
Table~\ref{tab:Pearson}, with column~4 showing the correlation
coefficients for the actual sample, to be contrasted with column~3,
that corresponds to a fully uncorrelated set of mock observations with
the same number of stars and with the same flux uncertainties
(see \S\ref{Ssec:Sims}).  The strong positive correlation between
A$_V$ and $T_{\rm eff}$ is confirmed, along with an anticorrelation
between NUV bump strength and A$_V$. Fig.~\ref{fig:DPars} shows that
the cooler subsample features a slight correlation between $\delta$
and A$_V$, and there is also a slightly positive correlation between
$\delta$ and $E_b$, so that steeper laws (i.e. lower values of
$\delta$) display weaker bumps.

Notice the trend between the steepness parameter
($\delta$) and A$_V$, with an overall value very close to the
commonly adopted Milky Way standard ($\delta$=$-$0.05, corresponding
to R$_V$=3.1). The cooler subsample features increasing $\delta$ with extinction. 
A simple interpretation of changes in
$\delta$ would be due to variations in the distribution of dust grain sizes,
with smaller grains producing steeper laws (i.e. tending to Rayleigh
scattering) and larger grains tending towards a weak wavelength
dependence (i.e. tending to Mie scattering). Our results 
suggest that the general distribution of dust particle sizes
does not depend strongly on the net amount of dust extinction, with
a weak correlation towards greyer extinction with increasing $A_V$
(expected in dust with a higher contribution from larger grains).
Note that the attenuation law in galaxies also becomes 
shallower at increasing amounts of
dust \citep[see, e.g.,][]{Chevallard:13,Hagen:17,Salim:18,Tress:18,Decleir:19},
an effect expected from the inhomogeneous distribution of
dust within the stellar populations of a galaxy \citep[e.g.,][]{WG:00}.

Fig.~\ref{fig:Aitoff} shows the spatial distribution of the dust
extinction parameters, from top to bottom: A$_V$, E$_b$ and
$\delta$. The Aitoff projections are shown in Galactic coordinates. We
should emphasize that our working sample originates 
from all available pointings made by the {\sl Swift}/UVOT
instrument, assembled in the Serendipitous Source Catalogue.
Therefore it is not a dedicated / programmed survey, and
only consists of what could be considered sets of random pointings on
the sky. This means that the sky is covered very sparsely, with many
sources located within relatively narrow patches -- the field of view of
UVOT is 17$\times$17\,arcmin$^2$ \citep{Roming:05}. Therefore, to create
a meaningful representation of the spatial variations of the dust parameters, 
we grouped all source detections within a single UVOT pointing and obtained
single measurements of (A$_V$, E$_b$, $\delta$) that
correspond to the median of the dust fits within the same
field of view. Within each group, we found that the scatter of the
dust parameters is roughly compatible with the expected uncertainties.
For instance, the scatter of the intra-group parameters 
-- i.e. the distribution of measurements of $A_V$, $E_b$, $\delta$ towards
stars within the same field of view of the UVOT camera -- has a median of
$\sigma(A_V)=0.15$, $\sigma(E_b)=1.7$, $\sigma(\delta)=0.15$.

The figure shows that most of our targets are located on a thin layer
of the Galactic plane -- we emphasize that this sample is constrained
to have all NUV fluxes with uncertainty below 20\%, therefore
restricting the sample to hotter, more luminous stars, mostly around
the turn-off point of the Main Sequence (see
Fig.~\ref{fig:Isoch}). The top panel illustrates the expected increase
in total extinction towards the plane of the Galaxy, more clearly
shown in Fig.~\ref{fig:ParSky}. The middle panel shows a random
distribution of NUV bump strengths. The
bottom panel shows some weak coherence in the variation of the
steepness parameter, $\delta$, not only in latitude but also in
longitude, with a preference for lower values of $\delta$ towards
$\ell\sim$30--120 degrees.

A more detailed view of these trends can be seen in
Fig.~\ref{fig:ParSky} where we show the projection of the dust
extinction parameters (vertical axes) along the three astrometric
quantities (from left to right: distance, Galactic latitude and
longitude), following the same running median criteria as in
Fig.~\ref{fig:DPars}.  In the top row it is worth highlighting the
strong increasing extinction towards low Galactic latitude, with
values A$_V\gtrsim$1 close to the Galactic plane ($|b|<10^{\rm o}$).
Note that this correlation confirms the validity of the parameter
constraints, as it is derived from totally independent methods
(i.e. astrometry vs flux ratios).  Moreover the strong anticorrelation
between A$_V$ and $|b|$ is found regardless of temperature, although
the cooler subsample extends to higher latitude, as shown in
Fig.~\ref{fig:Teff}. The trend between A$_V$ and distance depends on
temperature: at fixed distance, the hotter subsample features higher
extinction, a result due in part to the previous correlation and the
fact that hotter stars are preferentially found at lower Galactic
latitude (Fig.~\ref{fig:Teff}).  However, a selection bias might be
present, as the cooler stars at high extinction will not be included
in this NUV flux-limits sample.  The NUV bump strength (middle panels)
is highly covariant with distance, with the sightlines towards the
more distant stars showing the weaker bump strengths. Noting the 
trend between A$_V$ and E$_b$ shown in
Fig.~\ref{fig:DPars} along with the mild increase of A$_V$ with distance,
we could argue that the nearer stars probe regions of lower extinction, where the
bump is expected to be stronger. The strong
separation with respect to effective temperature is, once more,
apparent, as in Fig.~\ref{fig:DPars}.  Finally, in the bottom panels
we show $\delta$, that appears mostly correlated with latitude, with
steeper laws towards the Galactic plane. The previously mentioned
``dip'' at $\ell\sim$30--120 degrees is also evident here. It is also
worth mentioning that the cooler stars drive the trends of $\delta$,
possibly due to the fact that these stars are more  spread
in Galactic latitude, hence probing more diverse regions regarding
dust composition. This result would naively suggest that
at lower Galactic latitude, large dust grains are destroyed more
efficiently, leading to a steeper extinction law.
Note, though, that the behaviour of E$_b$ suggests no significant
variation with Galactic latitude in the cooler subset, and that A$_V$
{\sl decreases} quite strongly with latitude.  The interpretation of
these trends is non trivial given the interlinked dependence between
A$_V$, steepness and bump strength, requiring a detailed analysis
based on a dust model along with a proper radiative transfer treatment
in a disc-like geometry, beyond the scope of this paper.
We also note that no significant trends have been found in the other panels.

\section{Summary}
\label{Sec:Summ}

By cross-matching the {\sl Swift}/UVOT Serendipitous Source Catalogue
\citep{UVOTCat2} with Data Release 2 of the {\sl Gaia} mission \citep{GDR2}
and the {\sl 2MASS} point source catalogue \citep{2MASS}, we assembled
a sample of 10,452 stars with good photometry covering a wide
spectral range from the near ultraviolet (with filters that straddle
the well-known dust absorption feature at 2,175\AA ) to near infrared.
The sample extends out to several kpc (95\% of the sample is located within
3.7\,kpc), and is limited in effective temperature (T$_{\rm eff}>6,500$\,K)
in order to have enough flux in the NUV to constrain the NUV bump.
A comparison of the colours with a grid of synthetic stellar atmospheres
from the models of \citet{Coelho:14} allows us to constrain the most
fundamental parameters of dust extinction in this spectral window, namely the steepness of the
wavelength dependence ($\delta$), the strength of the NUV bump ($E_b$),
and the $V$ band extinction (A$_V$), following the
standard prescription of \citet{Noll:09}. We note that the serendipitous
nature of this data set produces sparse sampling on the sky, preventing
us from producing a dust map \citep[e.g.,][]{Chen:19, Lallement:19}.
However, this is the first time a large sample (comprising over 10
thousand stars) is used to constrain {\sl both} optical and NUV
features of the dust extinction law of the Galaxy.

In addition to the well-known correlation between total extinction and
Galactic latitude (Fig.~\ref{fig:ParSky}), we find in
Fig.~\ref{fig:DPars} that the NUV bump strength appears anticorrelated
with total extinction ($A_V$) and positively correlated with the
extinction law steepness ($\delta$). Comparisons with simulations
allow us to reject the hypothesis of a systematic from the inherent
degeneracies as the cause of these trends (Table~\ref{tab:Pearson}).
These trends are stronger in the subsample of hotter stars -- mostly
located towards low Galactic latitude.  The steepness of the
extinction law does not vary significantly when plotted with respect
to A$_V$, and appears consistent with the value adopted for the Milky
Way standard ($\delta=-0.05$).

In Fig.~\ref{fig:Aitoff}, we show the distribution of our sample in
Galactic coordinates, colour-coded according to the dust extinction
parameters. We note the sparsity of the set, as the parent sample
originates from the analysis of all sources found in archival images
of the {\sl Swift}/UVOT camera, and therefore does not follow an
optimal survey strategy. Despite the low coverage of the sky, the
strong trend between A$_V$ and Galactic latitude is apparent once
more. The steepness of the extinction law also appears
to be correlated with latitude, and perhaps longitude,
with a prominence of low $\delta$ (i.e. steeper law) around
$\ell\sim$30--120$^o$. A more detailed illustration of the trends
with astrometric quantities is shown in Fig.~\ref{fig:ParSky}.
This paper presents for the first time substantial correlations
between dust extinction parameters, analogous but not equivalent to
such variations in the dust {\sl attenuation} law in galaxies. A more
detailed analysis comprising large samples of spectroscopic data
including the NUV optical window \citep[e.g.][]{Gordon:09,Fitz:19} is needed to
confirm or refute the trends presented here.
Understanding in detail the extinction law in the Galaxy
is a very important stage in the more general area of galaxy
formation and evolution. Dust attenuation in galaxies remains
an open problem, where dust composition and its distribution among
the stellar populations play equally important roles
\cite[see, e.g.][and references therein]{Galliano:18}. Therefore,
constraints on dust extinction parameters derived from nearby
resolved systems allow us to understand one of these important
components. Our results emphasize that theoretical/numerical
models of galaxy formation have to take into account dust
composition variations to explain the observed trends in the
dust attenuation law, as simple models based on a fixed composition
would not be able to explain these trends in our own Galaxy.

\section*{Acknowledgements}
The referee is warmly thanked for a very careful reading of this paper, providing
useful comments and suggestions that
have improved this paper. This work has made use of data from the
European Space Agency (ESA) mission {\it Gaia} ({\tt
http://www.cosmos.esa.int/gaia}), processed by the {\it Gaia} Data
Processing and Analysis Consortium (DPAC, {\tt
http://www.cosmos.esa.int/web/gaia/dpac/consortium}). Funding for the
DPAC has been provided by national institutions, in particular the
institutions participating in the {\it Gaia} Multilateral Agreement.
This research made use of the cross-match service provided by CDS,
Strasbourg.  MP was supported by the UK Science and Technology
Facility Council (STFC), grant number ST/N000811/1. MP and PK
acknowledge support from the UK Space Agency.  The authors acknowledge
the use of the UCL Myriad High Throughput Computing Facility
(Myriad@UCL), and associated support services, in the completion of
this work.

\section*{Data availability}
This sample is extracted from a cross-match of publicly available
datasets (UVOT Serendipitous Source Catalogue, {\sl Gaia} DR2, and {\sl 2MASS}
point source catalogue). The
combined final dataset is available upon request.


\label{lastpage}


\begin{thebibliography}{}

\bibitem[\protect\citeauthoryear{Alzate, Bruzual, \& D{\'\i}az-Gonz{\'a}lez}{2021}]{Alzate:21}
Alzate J.~A., Bruzual G., D{\'\i}az-Gonz{\'a}lez D.~J., 2021, MNRAS, 501, 302

\bibitem[\protect\citeauthoryear{Bai et al.}{2019}]{GDR2_Teff}
Bai Y., Liu J., Bai Z., Wang S., Fan D., 2019, AJ, 158, 93

\bibitem[\protect\citeauthoryear{Bailer-Jones, et al.}{2018}]{GDR2_Dist}
Bailer-Jones C.~A.~L., Rybizki J., Fouesneau M., Mantelet G., Andrae R., 2018, AJ, 156, 58

\bibitem[\protect\citeauthoryear{Bradley et al.}{2005}]{Bradley:05}
Bradley J., Dai Z.~R., Erni R., Browning N., Graham G., Weber P.,
Smith J., et al., 2005, Sci, 307, 244

\bibitem[\protect\citeauthoryear{Bressan, et al.}{2012}]{Bressan:12}
Bressan A., Marigo P., Girardi L., Salasnich B., Dal Cero C., Rubele
S., Nanni A., 2012, MNRAS, 427, 127

\bibitem[\protect\citeauthoryear{Calzetti et al.}{2000}]{Calz:00}
Calzetti D., Armus L., Bohlin R.~C., Kinney A.~L., Koornneef J.,
Storchi-Bergmann T., 2000, ApJ, 533, 682

\bibitem[\protect\citeauthoryear{Cardelli et al.}{1989}]{CCM:89}
Cardelli J.~A., Clayton G.~C., Mathis J.~S., 1989, ApJ, 345, 245

\bibitem[\protect\citeauthoryear{Chen et al.}{2019}]{Chen:19}
Chen B.-Q., Huang Y., Yuan H.-B., Wang C., Fan D.-W., Xiang M.-S.,
Zhang H.-W., et al., 2019, MNRAS, 483, 4277

\bibitem[\protect\citeauthoryear{Chevallard et al.}{2013}]{Chevallard:13}
Chevallard J., Charlot S., Wandelt B., Wild V., 2013, MNRAS, 432, 2061

\bibitem[\protect\citeauthoryear{Coelho}{2014}]{Coelho:14}
Coelho P.~R.~T., 2014, MNRAS, 440, 1027

\bibitem[\protect\citeauthoryear{Coelho et al.}{2020}]{Coelho:20}
Coelho, P.~R.~T., Bruzual, G., Charlot, S., 2020, MNRAS, 491, 2025

\bibitem[\protect\citeauthoryear{Conroy, Schiminovich, \& Blanton}{2010}]{CSB10}
Conroy C., Schiminovich D., Blanton M.~R., 2010, ApJ, 718, 184 

\bibitem[\protect\citeauthoryear{Cutri et al.}{2003}]{2MASS}
Cutri R.~M., Skrutskie M.~F., van Dyk S., Beichman C.~A.,
Carpenter J.~M., Chester T., Cambresy L., et al., 2003, yCat, II/246

\bibitem[\protect\citeauthoryear{Decleir et al.}{2019}]{Decleir:19}
Decleir M., De Looze I., Boquien M., Baes M., Verstocken S., Calzetti
D., Ciesla L., et al., 2019, MNRAS, 486, 743

\bibitem[\protect\citeauthoryear{Draine}{2003}]{Draine:03}
Draine B.~T., 2003, ARA\&A, 41, 241

\bibitem[\protect\citeauthoryear{Duley \& Seahra}{1998}]{DS:98}
Duley W.~W., Seahra S., 1998, ApJ, 507, 874

\bibitem[\protect\citeauthoryear{Fischera \& Dopita}{2011}]{FD:11}
Fischera J., Dopita M., 2011, A\&A, 533, A117

\bibitem[\protect\citeauthoryear{Fitzpatrick \& Massa}{1990}]{FM:90}
Fitzpatrick E.~L., Massa D.~L., 1990, ApJS, 72, 163

\bibitem[\protect\citeauthoryear{Fitzpatrick}{1999}]{Fitz:99}
Fitzpatrick, E.~L., 1999, PASP, 111, 63

\bibitem[\protect\citeauthoryear{Fitzpatrick et al.}{2019}]{Fitz:19}
Fitzpatrick E.~L., Massa D., Gordon K.~D., Bohlin R., Clayton G.~C.,
2019, ApJ, 886, 108

\bibitem[\protect\citeauthoryear{Fitzpatrick \& Massa}{2007}]{FM:07}
Fitzpatrick, E.~L., Massa, D.~L., 2007, ApJ, 663, 320
 
\bibitem[\protect\citeauthoryear{Foreman-Mackey et al.}{2013}]{emcee}
Foreman-Mackey D., Hogg D.~W., Lang D., Goodman J., 2013, PASP, 125, 306 

\bibitem[\protect\citeauthoryear{Gaia Collaboration, et al.}{2018}]{GDR2}
Gaia Collaboration, et al., 2018, A\&A, 616, A1

\bibitem[\protect\citeauthoryear{Galliano, Galametz, \& Jones}{2018}]{Galliano:18}
Galliano F., Galametz M., Jones A.~P., 2018, ARA\&A, 56, 673

\bibitem[\protect\citeauthoryear{Gordon et al.}{2003}]{Gordon:03}
Gordon K.~D., Clayton G.~C., Misselt K.~A., Landolt A.~U., Wolff
M.~J., 2003, ApJ, 594, 279

\bibitem[\protect\citeauthoryear{Gordon, Cartledge, \& Clayton}{2009}]{Gordon:09}
Gordon K.~D., Cartledge S., Clayton G.~C., 2009, ApJ, 705,
1320

\bibitem[\protect\citeauthoryear{Gregg et al.}{2006}]{NGSL}
Gregg M.~D., Silva D., Rayner J., Worthey G., Valdes F., Pickles A.,
Rose J., et al., 2006, hstc.conf, 209

\bibitem[\protect\citeauthoryear{Hagen et al.}{2017}]{Hagen:17}
Hagen L.~M.~Z., Siegel M.~H., Hoversten E.~A., Gronwall C., Immler S.,
Hagen A., 2017, MNRAS, 466, 4540

\bibitem[\protect\citeauthoryear{Hoversten et al.}{2011}]{Hoversten:11}
Hoversten E.~A., Gronwall C., Vanden Berk D.~E., Basu-Zych A.~R.,
Breeveld A.~A., Brown P.~J., Kuin N.~P.~M., et al., 2011, AJ, 141, 205

\bibitem[\protect\citeauthoryear{Hutton et al.}{2014}]{Hutton:14}
Hutton S., Ferreras I., Wu K., Kuin P., Breeveld A., Yershov V.,
Cropper M., et al., 2014, MNRAS, 440, 150

\bibitem[\protect\citeauthoryear{Hutton et al.}{2015}]{Hutton:15}
Hutton, S., Ferreras, I., Yershov, V., 2015, MNRAS 452, 1412

\bibitem[\protect\citeauthoryear{Kriek \& Conroy}{2013}]{KC:13}
Kriek, M., Conroy, C., 2013, ApJ, 775, L16      

\bibitem[\protect\citeauthoryear{Lallement et al.}{2019}]{Lallement:19}
Lallement R., Babusiaux C., Vergely J.~L., Katz D., Arenou F., Valette
B., Hottier C., et al., 2019, A\&A, 625, A135 

\bibitem[\protect\citeauthoryear{Narayanan et al.}{2018}]{Nara:18}
Narayanan D., Conroy C., Dav{\'e} R., Johnson B.~D., Popping G., 2018, ApJ, 869, 70

\bibitem[\protect\citeauthoryear{Noll et al.}{2009}]{Noll:09}
Noll, S.,  et al., 2009, A\&A, 499, 69

\bibitem[\protect\citeauthoryear{Oke \& Gunn}{1983}]{ABmag}
Oke J.~B., Gunn J.~E., 1983, ApJ, 266, 713

\bibitem[\protect\citeauthoryear{Panuzzo et al.}{2007}]{Panuzzo:07}
Panuzzo, P., Granato, G.~L., Buat, V., Inoue, A.~K., Silva, L., Iglesias-P\'aramo, J.,
Bressan, A., 2007, MNRAS, 375, 640

\bibitem[\protect\citeauthoryear{Pei}{1992}]{Pei:92}
Pei, Y.~C., 1992, ApJ, 395, 130

\bibitem[\protect\citeauthoryear{Page et al.}{2014}]{UVOTCat1}
Page M.~J., et al., 2014, Proceedings of Scinec (SWIFT-10), 037

\bibitem[\protect\citeauthoryear{Roming et al.}{2005}]{Roming:05}
Roming P.~W.~A., Kennedy T.~E., Mason K.~O., Nousek J.~A., Ahr L.,
Bingham R.~E., Broos P.~S., et al., 2005, SSRv, 120, 95

\bibitem[\protect\citeauthoryear{Salim, Boquien, \& Lee}{2018}]{Salim:18}
Salim S., Boquien M., Lee J.~C., 2018, ApJ, 859,
11

\bibitem[\protect\citeauthoryear{Sparke \& Gallagher}{2007}]{SG:07}
Sparke L.~S., Gallagher J.~S., 2007, gitu.book

\bibitem[\protect\citeauthoryear{Tress, et al.}{2018}]{Tress:18}
Tress M., et al., 2018, MNRAS, 475, 2363

\bibitem[\protect\citeauthoryear{Tress, et al.}{2019}]{Tress:19}
Tress M., et al., 2019, MNRAS, 488, 2301

\bibitem[\protect\citeauthoryear{Valencic, Clayton, \& Gordon}{2004}]{Valencic:04}
Valencic L.~A., Clayton G.~C., Gordon K.~D., 2004, ApJ, 616, 912


\bibitem[\protect\citeauthoryear{Weingartner \& Draine}{2001}]{Wein:01}
Weingartner J.~C., Draine B.~T., 2001, ApJ, 548, 296

\bibitem[\protect\citeauthoryear{Witt \& Gordon}{2000}]{WG:00}
Witt A.~N., Gordon K.~D., 2000, ApJ, 528, 799

\bibitem[\protect\citeauthoryear{Yershov et al.}{2014}]{UVOTCat2}
Yershov, V., et al., 2014, Astrophysics \& Space Science, 354, 97



\end{thebibliography}
\end{document}